\documentclass{aastex}
\usepackage{natbib}
\usepackage{emulateapj5}

\begin{document}

\title{The Century Survey Galactic Halo Project I:  Stellar Spectral Analysis}

\author{Warren R.\ Brown\altaffilmark{1}}

\affil{Harvard-Smithsonian Center for Astrophysics, 60 Garden St,
Cambridge, MA 02138}
\email{wbrown@cfa.harvard.edU}

\author{Carlos Allende Prieto}

\affil{McDonald Observatory and Department of Astronomy, 
University of Texas, Austin, TX 78712}

\author{Timothy C.\ Beers}

\affil{Department of Physics and Astronomy, Michigan State University,
East Lansing, MI 48824}

\author{Ronald Wilhelm}

\affil{Department of Physics, Texas Tech University, Lubbock, TX 79409}

\author{Margaret J.\ Geller,
	Scott J.\ Kenyon\altaffilmark{1},
	\and
	Michael J.\ Kurtz}

\affil{Smithsonian Astrophysical Observatory, 60 Garden St, Cambridge, MA
02138}

\altaffiltext{1}{Visiting Astronomer, Kitt Peak National Observatory,
National Optical Astronomy Observatories, which is operated by the
Association of Universities for Research in Astronomy, Inc. under
cooperative agreement with the National Science Foundation.}

\shorttitle{Century Survey Galactic Halo Project I}
\shortauthors{Brown et al.}

\begin{abstract}

The Century Survey Galactic Halo Project is a photometric and
spectroscopic survey from which we select relatively blue stars
($\vr<0.30$ mag) as probes of the Milky Way halo.  The Survey strip spans
the range of Galactic latitude $35^\circ<b<88^\circ$, allowing us to study
the nature of populations of stars and their systematic motions as a
function of Galactic latitude. One of our primary goals is to use blue
horizontal-branch stars to trace potential star streams in the halo, and
to test the hierarchical model for the formation of the Galaxy.

In this paper we discuss spectroscopy and multi-passband photometry for a
sample of 764 blue stars in the Century Survey region.  Our sample
consists predominantly of A- and F-type stars.  We describe our techniques
for determination of radial velocities, effective temperatures,
metallicities, and surface gravities.  Based on these measurements, we
derive distance estimates by comparison with a set of calibrated
isochrones.  We devote special attention to the classification of blue
horizontal-branch stars, and compare the results obtained from the
application of the techniques of Kinman et al., Wilhelm et al., and
Clewley et al.  We identify 55 blue horizontal-branch stars.  Our large
sample of stars also uncovers a number of unusual objects, including three
carbon-enhanced stars, a late B-type star located 0.8 kpc above the
Galactic plane, and a DZ white dwarf.

\end{abstract}


\section{INTRODUCTION}

The spatial distribution, motion, and composition of halo stars provide
a record of the Milky Way's past.  A wealth of surveys, including proper
motion- and metallicity-selected surveys of solar-neighborhood stars,
surveys of globular clusters and distant giant stars, and pencil-beam
surveys of dwarf stars have probed the structure of the halo \citep[e.g.,
see review by][]{majewski93}.  These surveys have explored the formation
of the halo, the thick disk, and the thin disk, often with dissenting
conclusions.

A photometric survey has an advantage over more traditional surveys
because there are no selection biases in velocity, proper motion, or
metallicity for halo and thick-disk stars.  Recent work \citep{ivezic00,
yanny00, vivas01, newberg02, vivas03} shows that photometric surveys of RR
Lyrae and blue horizontal branch stars can reach very deep, and can
identify structure in the halo at distances of $\sim$100 kpc.  However,
these photometric surveys sacrifice both the full 6-dimensional kinematic
information provided by radial velocities and proper motions and the
abundance information that can be obtained from a spectroscopic study.  
Here we discuss the first results from the Century Survey Galactic Halo
Project, a photometric and spectroscopic survey of color-selected stars in
the halo and thick disk of the Galaxy.

The Century Survey is a galaxy redshift survey \citep{geller97} for which
we obtained 64 deg$^2$ of $V$ and $R$ imaging to measure a multi-passband
galaxy luminosity function \citep{brown01}.  Here we use this CCD
photometry to select blue $(\vr)<0.30$ mag stars for follow-up
spectroscopy.  Moderate signal-to-noise (S/N$\approx$30) spectra allow
us to measure radial velocities, temperatures, surface gravities, and
metallicities for the stars, with the goal of probing the nature and the
structure of the Milky Way halo and thick-disk populations.

Previous surveys demonstrate that blue horizontal branch (BHB) stars
provide an excellent probe of the halo \citep{pier82, sommer89, preston91,
arnold92, kinman94, wilhelm99b}.  One advantage of using BHB stars as
tracers is that they are numerous, exceeding the abundance of RR Lyraes by
roughly a factor of 10 \citep{preston91}.  Another advantage is that BHB
stars are relatively luminous and hence observable to large distances.  
Furthermore, BHB stars have a small dispersion in absolute magnitude, which
makes precise distance estimates possible.

A major difficulty in using BHB stars as probes of Galactic structure is
the need to reliably distinguish between low surface-gravity BHB stars and
the higher surface-gravity A-type dwarfs and blue stragglers.  BHB stars
are core helium-burning stars with lower surface gravities than
main-sequence stars of the same spectral type.  Although investigators
once thought blue stragglers were a minor component of the halo
population, recent studies \citep{norris91, preston94, wilhelm99b}
demonstrate that a surprisingly large fraction of faint stars in the color
range associated with BHB stars are indeed high-gravity stars, many of
which are blue stragglers \citep[see][]{preston00}. Distinguishing BHB
stars is particularly important for our blue star sample, which contains a
large number of A dwarfs, F dwarfs, and some subgiants.  The A and F
dwarfs probe the thick and thin disk; the BHB stars probe the inner halo.  
To distinguish between BHB and A/F dwarfs, we investigate the surface
gravity measures of \citet{kinman94}, \citet{wilhelm99a}, and
\citet{clewley02}.  We compare the results of these three methods, and
make our BHB selection based on this comparison.

The $1^\circ\times64^\circ$ Century Survey photometric strip is
located at $8\fh5 < \alpha_{1950} < 13\fh5, 29^\circ < \delta_{1950} <
30^\circ$. In Galactic coordinates the Century Survey strip cuts across
$35^\circ \le b \le 85^\circ$, along a line of constant Galactic longitude
$l \approx 200^\circ$, before crossing near the north Galactic pole and
dropping to $b=80^\circ$ along $l \approx 50^\circ$ (see Figure
\ref{fig:skymap}). The placement and depth of the Century Survey
photometry allows us to address a number of important science goals.

Recent observations and n-body simulations lend increasing support to a
hierarchical picture where the halo of the Galaxy is composed (at least
partially) of tidally disrupted dwarf galaxies \citep[e.g.][]{searle78}. A
good example is the discovery of the Sagittarius dwarf galaxy in the
process of tidal disruption by the Milky Way \citep{ibata94}. N-body
models suggest that dwarf galaxies disrupted long ago should still be
visible as streams of stars within the Galaxy's halo \citep{johnston96}.
If the halo potential is spherically symmetric \citep{ibata01}, a
90$^\circ$ strip will, in principle, sample 1/2 of every star stream
orbiting the Galaxy, and hence strongly constrain the merger history of
the Galactic halo. Thus the Century Survey Galactic Halo Project is well
suited to testing the hierarchical picture for the formation of the Milky
Way.

The Century Survey Galactic Halo Project strip spans a wide range of
Galactic latitude on both sides of the north Galactic pole and can provide
a robust picture of the systematic motions of the thick disk and halo.
There is already evidence for systematic motions at all scales in the
Milky Way halo. For example, \citet{kinman96} found 24 BHB and RR Lyrae
stars (out of a sample of 69 stars) streaming towards us from the north
Galactic pole with velocity $-59\pm16$ km s$^{-1}$. \citet{majewski96}
observed large-scale retrograde motion in a deep proper motion sample of
250 halo stars at the north Galactic pole. \citet{gilmore02} recently
reported evidence for a surprisingly low mean rotational velocity for at
least a portion of the thick-disk population, possibly associated with
debris from an ancient satellite merger. All of these results are based on
surveys which cover only a few square degrees, hence a coherent picture of
halo/thick-disk motion is difficult to obtain. As an example of what might
be gained from larger area surveys, \citet{yanny03} use the Sloan Digital
Sky Survey to suggest the presence of a ``ring'' of stars close to the
plane of the Milky Way. This ``ring'' might be associated with a tidally
disrupted satellite.  Interestingly, the bright ``metal-weak thick-disk''
stars from the study of \citet{beers02} exhibit a similar range of
metallicity to those inferred for the SDSS ring stars, and kinematics that
are similar to the more metal-rich stars of the suggested ``intruder
population'' described by \citet{gilmore02}.

The placement of the Century Survey photometric region also allows us to
establish well-defined, magnitude-limited samples of thin disk/thick disk/halo
A-type stars as a function of Galactic latitude \citep{rodgers71,lance88a,
rodgers93}. We will also find distant, high-latitude OB main-sequence
stars \citep{brown89,conlon90}.

In this paper we discuss the spectroscopic analysis of 764 blue stars from
the on-going Century Survey Galactic Halo Project. In \S \ref{sec:sample}
we describe our photometry, spectroscopy, and the efficacy of our sample
selection in finding A-type stars. In \S \ref{sec:vel} we discuss our
determinations of radial velocities. In \S \ref{sec:pars} we describe our
methodology for estimation of stellar effective temperatures, surface
gravities, and metallicities. In \S \ref{sec:class} we describe our
stellar classification, and in \S \ref{sec:dist} we describe our distance
estimates. We devote special attention to BHB classification methods in \S
\ref{sec:bhb}. In \S \ref{sec:results} we describe the properties of our
sample and list some unusual stellar objects. Our conclusions are
presented in \S \ref{sec:concl}.

\section{THE SAMPLE} \label{sec:sample}

\subsection{Photometry}

The Century Survey photometry is based on Johnson $V$ and Cousins $R$
broadband imaging obtained with the 8 CCD MOSAIC camera \citep{muller98}
on the KPNO 0.9 m telescope in 1998 December and 1999 February.  The
imaging covers a $1.0^\circ\times64.0^\circ$ strip located at $8^{\rm
h}32^{\rm m}45^{\rm s} < \alpha_{\rm B1950} < 13^{\rm h}27^{\rm m}31^{\rm
s}$, $29^\circ < \delta_{\rm B1950} < 30^\circ$.  \citet{brown01} includes
a detailed description of the data reduction.  In brief, the astrometric
and photometric solutions are accurate to $\pm0\farcs4$ and $\pm0.03$ mag,
respectively.  The average depth of the photometry is $V=20.3$ mag.  The
photometric errors are dominated by the zero-point errors at bright
magnitudes, resulting in an average color accuracy of
$\sigma{(\vr)}=\pm0.042$ mag for the $V<16.5$ mag Century Survey Galactic
Halo Project sample selected for spectroscopy.

We obtained $JHK$ photometry from the Two Micron All Sky Survey (2MASS)
second incremental data release \citep{skrutskie00} for $J<15$ mag stars
in the Century Survey region as well as for stars in a
$1^\circ\times65.33^\circ$ region adjacent to the Century Survey.  The
adjacent 2MASS region is used to expand the area of the Century Survey
Galactic Halo Project.  We broke the adjacent region into two pieces to
match the available 2MASS photometry:  $8\fh5 < \alpha_{\rm B1950} <
11\fh25$, $28^\circ < \delta_{\rm B1950}< 29^\circ$ and $11\fh25 <
\alpha_{\rm B1950} < 13\fh5$, $30^\circ < \delta_{\rm B1950}< 31^\circ$.  
Photometry was available for 58.03 deg$^2$, or 88.8\%, of the adjacent

 \includegraphics[width=3.3in, bb=35 230 565 525,clip=true]{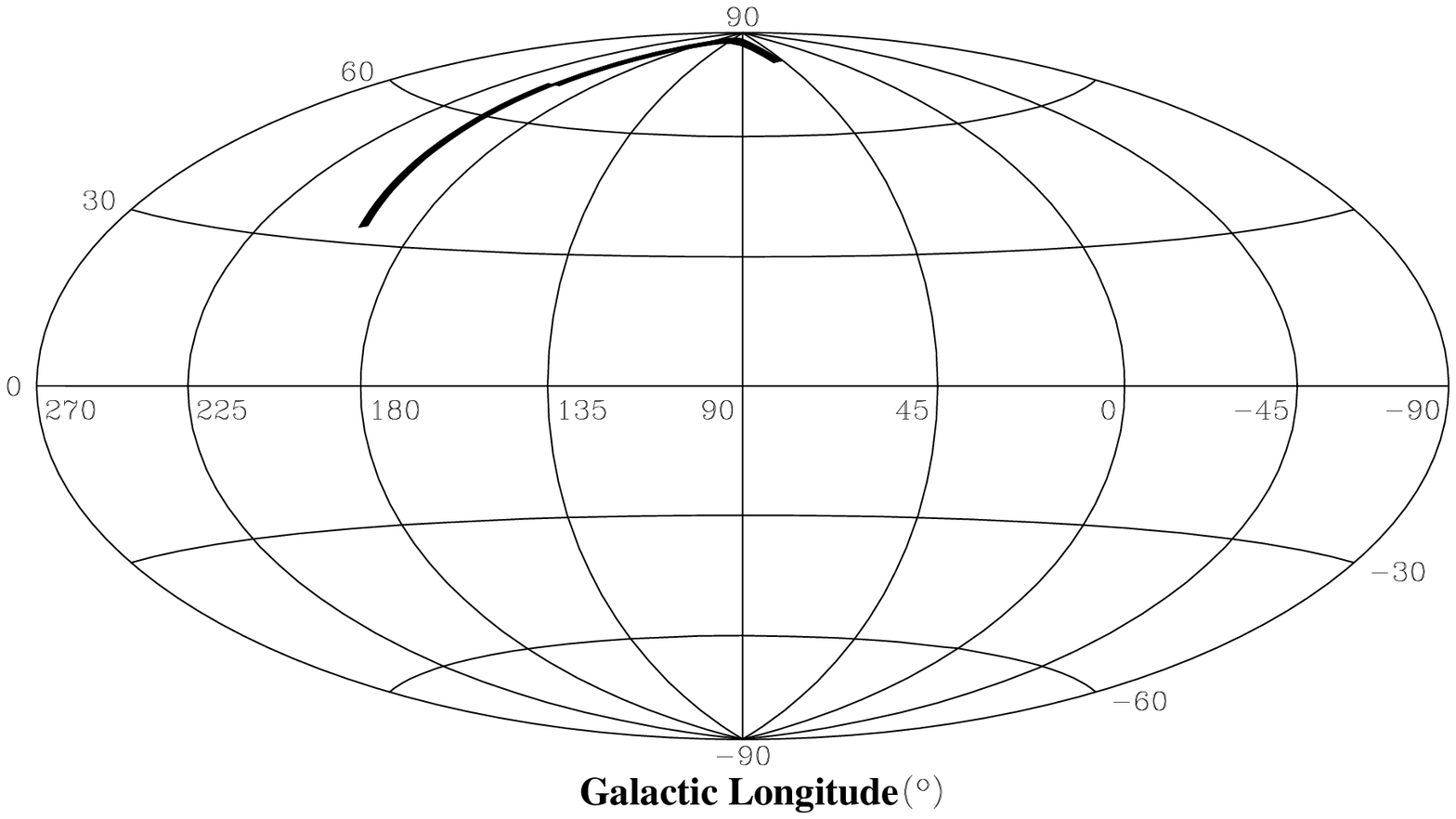}
 \figcaption{ \label{fig:skymap}
	Placement of the Century Survey and 2MASS photometric regions in
Galactic coordinates.  For display purposes we have placed +90$^\circ$
longitude, the direction of the solar orbit around the Galaxy, at the
origin.}

	~ \\

\noindent 2MASS region.  At spectral type A0, the depth of the $J<15$
mag adjacent 2MASS region is equivalent to $V<15$ mag.  The average
$(J-H)$ error at $J=15$ mag is $\sigma{(J-H)}=\pm0.09$ mag.

Figure \ref{fig:skymap} shows the placement of the photometric regions in
Galactic coordinates.  The Century Survey photometry cuts across $35^\circ
< b < 85^\circ$ along a line of constant Galactic longitude $l \approx
200^\circ$ (the direction of the Galactic anti-center) before crossing
near the north Galactic pole at $b=88^\circ$ and dropping to $b=80^\circ$
at $l \approx 50^\circ$.  Note that, for display purposes, we have placed
$l=90^\circ$ at the center of Figure \ref{fig:skymap}.

\subsection{Photometric Selection}
	
Table \ref{tab:obs} summarizes the selection of the 764 blue-star
candidates. We selected the primary sample of stars from the
$1^\circ\times64^\circ$ Century Survey photometry region with $V<16.5$ mag
and $(\vr)<0.25$ mag. We also obtained spectra for the redder, brighter
($0.25\le(\vr)<0.30$ mag ; $V<15.5$ mag) stars during the Spring 2001
observing season. Because the redder stars contained no BHB candidates and
outnumbered the $(\vr)<0.25$ mag stars by a factor of 2.4, we observed
1-in-10 of the redder stars in the $V<16.5$ mag sample during the Spring
2002 observing season. The blue $(\vr)<0.25$ mag, $V<16.5$ mag sample
selection yields 45\% A-type stars and 44\% F-type stars.

One hundred eleven of the 764 stars were selected with $J<15.0$ mag and
$(J-H)<0.15$ mag from the 2MASS region adjacent to the Century Survey. The
2MASS second incremental data release photometric selection yields 51\%
A-type stars and 38\% F-type stars.

\subsection{Spectroscopy}

We obtained medium-resolution spectra for our sample of 764 blue-star
candidates with the FAST spectrograph \citep{fabricant98} on the Whipple
Observatory 1.5 m telescope, during the Spring 2001 and Spring 2002
observing seasons.  We use a 600 line mm$^{-1}$ grating and a 2 arcsec
slit to obtain a resolution of 2.3 \AA\ and a spectral coverage from 3400
to 5400 \AA. Our exposure times vary; they are designed to reach a
signal-to-noise (S/N) ratio of 30 at 4000 \AA. We allow S/N=15 for objects
at $V=16.5$ mag.


The spectra are processed in the usual way with IRAF\footnote{IRAF is
distributed by the National Optical Astronomy Observatories, which are
operated by the Association of Universities for Research in Astronomy,
Inc., under cooperative agreement with the National Science Foundation.}.  
We first subtract a nightly bias frame from the raw images, but only use a
dark frame when the dark current exceeds 1 count per pixel for our
exposure times. (The dark current was atypically high for about a month
following the UV-illumination of the CCD, a back-side illuminated Loral
chip.)  We then create a normalization frame by dividing a high-order
cubic spline fit from the nightly flat fields.  We smooth the bluest
200 \AA\ of the normalization frame with a 3 pixel box to reduce
pixel-to-pixel noise in the lower S/N blue end of the flat field.

We extract one-dimensional spectra with the IRAF $apextract$ package.  
Wavelength calibrations are determined from helium-argon lamp comparison
spectra taken immediately after each observation.  The wavelength
solutions use a 3rd order polynomial fit of $\sim$40 spectral lines, with
RMS residuals of $\pm0.07$ \AA. The spectra are then flux calibrated with
nightly standard star observations, usually of Feige 34 or HZ 44
\citep{massey88}. The accuracy of the flux calibration is limited by the
fact that, for observational efficiency, we do not rotate the slit to the
parallactic angle.  However, the objects are well placed in the sky at
Mt.\ Hopkins:  we observe 68\% below an airmass of 1.15.  The absolute
flux calibration is good to 10\% for objects observed in photometric
conditions; 40\% of the objects were observed through light cirrus or in
poor seeing and do not have an accurate absolute flux calibration.

\section{RADIAL VELOCITIES} \label{sec:vel}

We measure stellar radial velocities in two ways, (1) with the
cross-correlation package RVSAO \citep{kurtz98} and (2) by measuring the
central wavelengths of several strong lines and comparing them to their
rest values.  We find an average offset of $-0.3\pm10.4$ km s$^{-1}$
between the two methods.  The dispersion is consistent with our external
error of $\pm10$ km s$^{-1}$. Our final velocities are an average of the
cross-correlation and line-by-line approach.

\subsection{Cross-Correlation}

To implement this approach, we constructed cross-correlation templates
from observations of 36 radial-velocity standards.  These standards have
known velocities accurate to $\pm0.01$ km s$^{-1}$
\citep{fekel99,stefanik99,udry99}. We also make use of observations of 25
bright spectral-type standards \citep{jacoby84}; the velocities for these
stars are from \citet{wilson63}, and have a typical accuracy of $\pm2$ km
s$^{-1}$.

It is important to have a range of cross-correlation templates because,
for a particular observation, a template with a significantly different
spectral type produces an asymmetry in the cross-correlation peak and a
systematic offset in the measured velocity.  We thus bin the observations
by spectral sub-type to make 16 templates spanning from B through early K.  
Except for the B templates, there are four or more stars averaged together
to construct a template.

The RVSAO package normalizes the 3700--5400 \AA\ region of our
spectra for cross-correlation.  We find the highest correlation peaks when
we multiply the Fourier amplitudes with a cosine-bell filter starting at
20 pixels and running to 1024 pixels.  The average internal error from the
cross-correlation is $\pm3.5$ km s$^{-1}$.

We measure our external velocity error by comparing 23 BHB star velocities
with velocities published by \citet{kinman94}.  We find a mean +3.5 km
s$^{-1}$ offset and a $\pm19.2$ km s$^{-1}$ RMS dispersion relative to the
\citet{kinman94} velocities.  Subtracting the published error $\sigma_{\rm
Kinman}=\pm16.2$ km s$^{-1}$ from the measured dispersion of $\sigma_{\rm
measured}=\pm19.2$ km s$^{-1}$ leaves us with an external error of
$\pm10.3$ km s$^{-1}$.  This external error is comparable with the
dispersion of the individual standard star observations.  Standard stars
were observed 3-5 times and had an RMS dispersion of $\pm9.7$ km s$^{-1}$.

We test the effects of decreasing S/N on our cross-correlation
measurements by adding Gaussian noise to high-S/N standards.  For
S/N=30 at 4000 \AA\ the RMS dispersion in velocity is $\pm6$ km s$^{-1}$;
thus the stars with $V<15.5$ mag have radial velocity errors dominated by
our external error of $\pm10$ km s$^{-1}$.  For S/N=15 at 4000 \AA\ the
RMS dispersion in velocity increases to $\pm15$ km s$^{-1}$; thus stars at
our $V=16.5$ mag limit have total radial velocity errors approaching
$\pm20$ km s$^{-1}$.

As a further check, we have independently obtained cross-correlation
velocities by employing a set of 16 synthetic stellar templates (with, by
definition, zero velocity and infinite S/N ratios) covering a range of
temperatures, gravities, and metallicities, constructed as described in
\citet{wilhelm99a}.  In all cases, the measured cross-correlation
velocities obtained by this method are consistent, within the expected
errors, with those obtained from the approach above.

\subsection{The ``Line-by-line'' Approach }

The line-by-line method optimally locates \citep[using the Gaussian
derivative technique described in detail in][]{beers90} the centers of
prominent absorption lines in the spectra (e.g.\ Ca {\small II} K,
H-$\delta$, Ca {\small I} 4226 \AA, H-$\gamma$, H-$\beta$), and obtains an
averaged radial velocity after pruning of discrepant lines.  This method
provides a valuable complement to the cross-correlation approach, both as
a reality check, and for cases (such as the A-type stars) where the
breadth of the Balmer lines results in a ``soft'' peak in the
cross-correlation function.  This method also enables extraction of radial
velocities for stars that are not well-matched by the range of templates
we use in the cross-correlations.  Internal errors in this approach are
roughly 7-10 km s$^{-1}$.  Extensive tests of the line-by-line method
during the course of the HK survey of Beers and colleagues indicate that
the external errors are similar.

Appendix Data Table \ref{tab:dat2} lists the final adopted radial
velocities for our program objects.  Given our precision, we ignore some
of the caveats involved in the definition of radial velocity, but refer
the reader to the interesting discussion in \citet{lindegren03}.

\section{STELLAR PARAMETERS} \label{sec:pars}

The derivation of physical parameters from our observed spectra is
basically an optimization problem with a unique solution for effective
temperature $T_{\rm eff}$, surface gravity $\log{g}$, and ``metallicity,''
which we assume to be proportional to the iron abundance, [Fe/H]. We use
libraries of synthetic spectra, and search for the set of these parameters
that best reproduces a given observation with a genetic algorithm. We
compare our results against the independent methods of \citet{beers99} and
\citet{wilhelm99a}. We estimate uncertainties in $T_{\rm eff}$, $\log{g}$,
and [Fe/H] by comparing the values from all three methods.

\subsection{Genetic Algorithm}

The spectral range of the Century Survey Galactic Halo Project
observations is sufficiently large to provide several indicators of
stellar surface temperature.  The slope of the observed continuum is
mainly dependent on the atmospheric $T_{\rm eff}$.  However, the modest
accuracy of the flux calibration for the majority of our program stars
limits its usefulness.  The Balmer-line profiles are also very sensitive
to $T_{\rm eff}$, but for earlier spectral types and lower metallicities
the lines are significantly affected by gravity and metallicity. In
addition, the Balmer lines are difficult to model because their formation
depend on (poorly understood) convective energy transport in stellar
envelopes.

Fortunately, the wavelength coverage of our spectra extends to include the
blue side of the Balmer jump.  The Balmer jump not only provides a
well-understood gravity indicator, but also serves to decouple the effects
of gravity from temperature on the damping wings of the Balmer lines.  At
a resolving power of $R \sim 2000$, several strong metal lines react
mainly to $T_{\rm eff}$ and the metallicity.  The Ca {\small II} K line is
typically saturated at solar metallicity, but it remains as the only
reliable metallicity sensitive feature for [Fe/H] $\lesssim -2$.  For the
warmest and most metal-poor stars in our sample, especially at low S/N, we
cannot establish all three parameters with certainty; the sensitivity to
metallicity is the first to be lost.

We compare our observed spectra with the low-resolution spectra calculated
by \citet{kurucz93}, based on plane-parallel LTE line-blanketed model
atmospheres between 3500 and 5300 \AA. We also compute a grid of synthetic
spectra with the code {\it Synspec} \citep{hubeny00}, using the same
model atmospheres. We compute synthetic spectra with a resolution matching
our observations in two windows: 3810--4010 \AA\ and 4700--5000 \AA. We
use very simple continuum opacities (H, H$^{-}$, electron and Rayleigh
scattering). The effect of metals in the continuum is negligible for our
combination of spectral range, stellar parameters, and S/N. The 4700--5000
\AA\ window, centered around H$\beta$, is very effective for stellar
classification of F--K stars with solar metallicity and similar-type
moderately metal-poor stars \citep{allende03}. The 3810--4010 \AA\ window,
which includes the the Ca {\small II} H and K lines as well as the
higher-order members of the Balmer series, is useful for extracting
physical parameters for the metal-poor A--F stars in our survey.


To reproduce the collisionally enhanced wings of the Ca {\small II} lines
we adopted damping parameters from \citet{barklem00}.  We also modeled the
Balmer lines with new calculations of the absorption coefficients as in
\citet{barklem02}.  We vary the micro-turbulence $\xi$ in the calculation
of the $R\simeq 2000$ spectra; its value is fixed at 2 km s$^{-1}$ in the
low-resolution spectra computed by Kurucz.  The synthetic spectra cover
the following range in the stellar parameters:
	\begin{eqnarray}
\begin{tabular}{cccccc}
4500 & $\le$ & $T_{\rm eff}$ & $\le$ & 10,000 & K\\
2.0 & $\le$ & $ \log g$ & $\le$ & 5.0 & dex \\
$-4.5$ & $\le$ & [Fe/H] & $\le$ &  $+0.5$ & dex \\
0 & $\le$ &  $\xi$ & $\le$ &  2 & km s$^{-1}$. \\
\end{tabular}
	\end{eqnarray} We assume that the [Mg/Fe] and [Ca/Fe] ratios are solar
for [Fe/H]$\ge 0$, +0.4 for [Fe/H]$< -1.5$, and vary linearly with declining
[Fe/H] between those two ranges.

We use a genetic algorithm \citep[GA;][]{carroll01} to search over the
parameter space and to find the optimal match for each star.  The final
parameters we adopt are the average of runs using a micro-GA with uniform
cross-over and a regular GA with creep mutation.  We use multilinear
interpolation to transform the discrete grid of synthetic spectra into a
continuous function for the GA.  Figure \ref{fitting} compares observed
and best-matching model spectra for four stars in the sample.  Because of
the increasing noise, the Balmer lines in the left window have a very low
weight in comparison with H$\beta$. We also assign lower weights to the
core of strong (H and Ca {\small II}) lines, which are poorly reproduced
because of expected departures from LTE in high atmospheric layers.  Our
metallicity determination for warm and/or metal-poor stars relies on the
strength of the Ca {\small II} K line and, therefore, we may have large
systematic errors for the possible low-metallicity $\alpha$-poor stars in
our sample \citep{carney97,king97}.

 \includegraphics[width=2.5in, angle=90, clip=true]{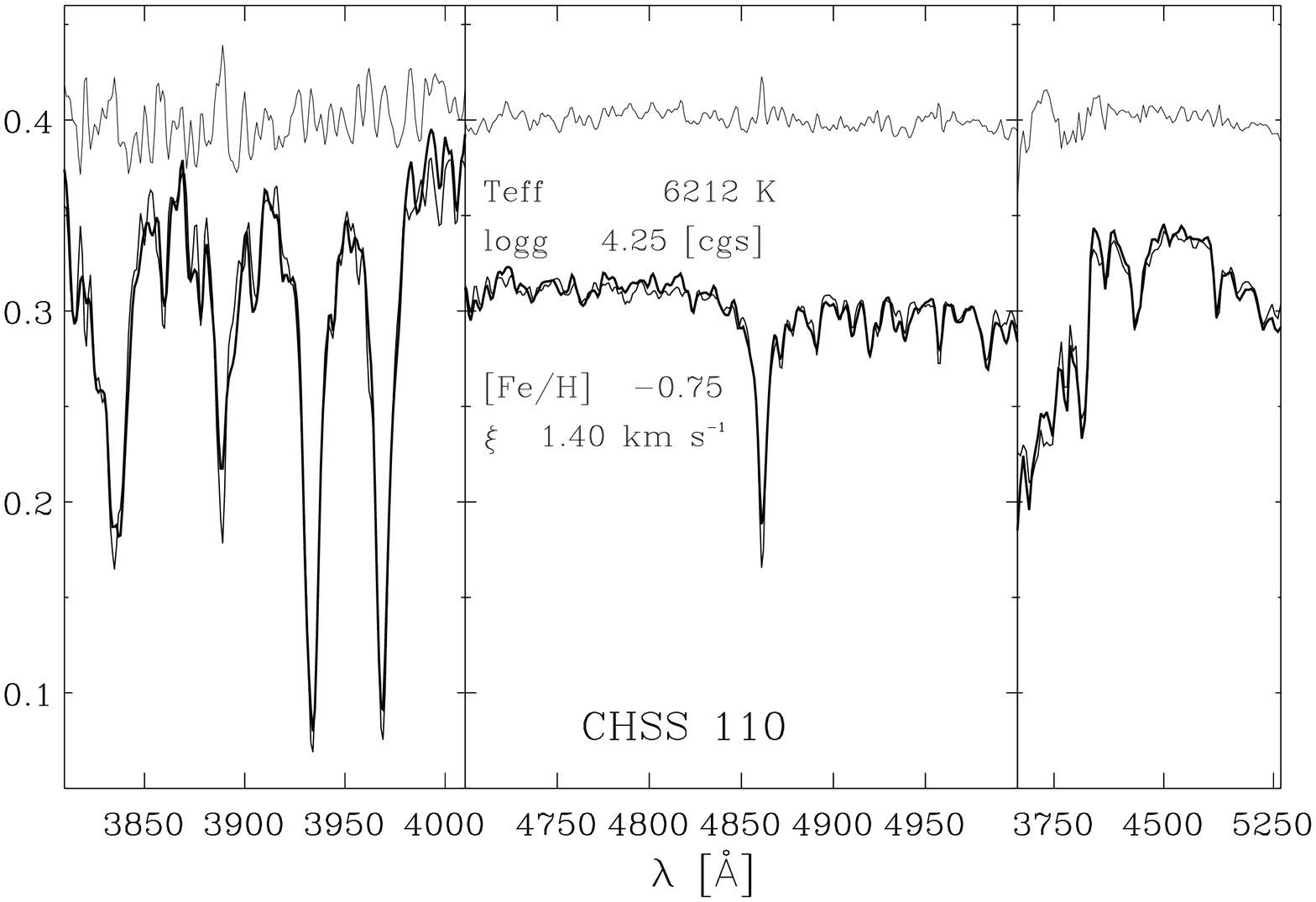}
  \includegraphics[width=2.5in, angle=90, clip=true]{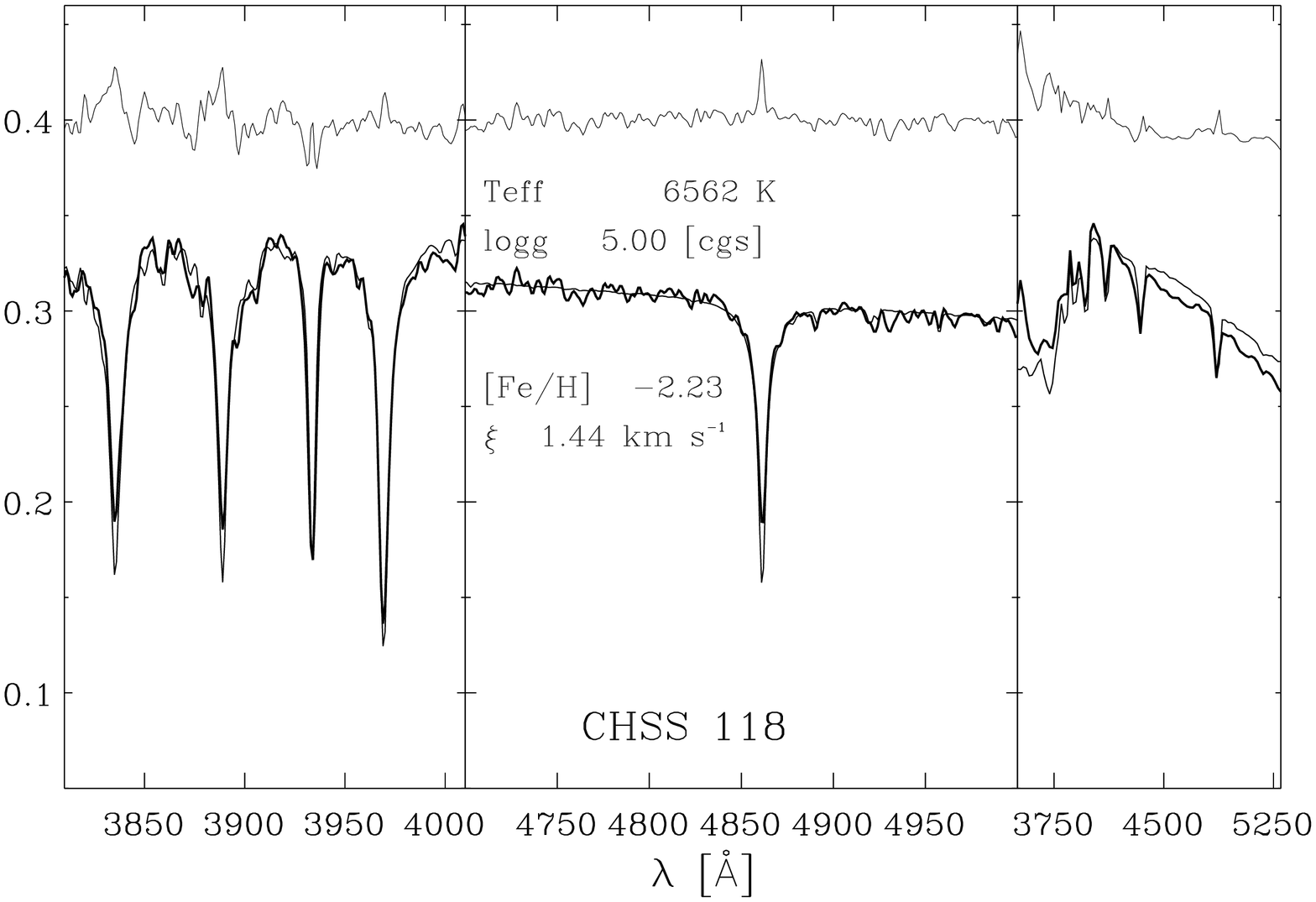}
   \includegraphics[width=2.5in, angle=90, clip=true]{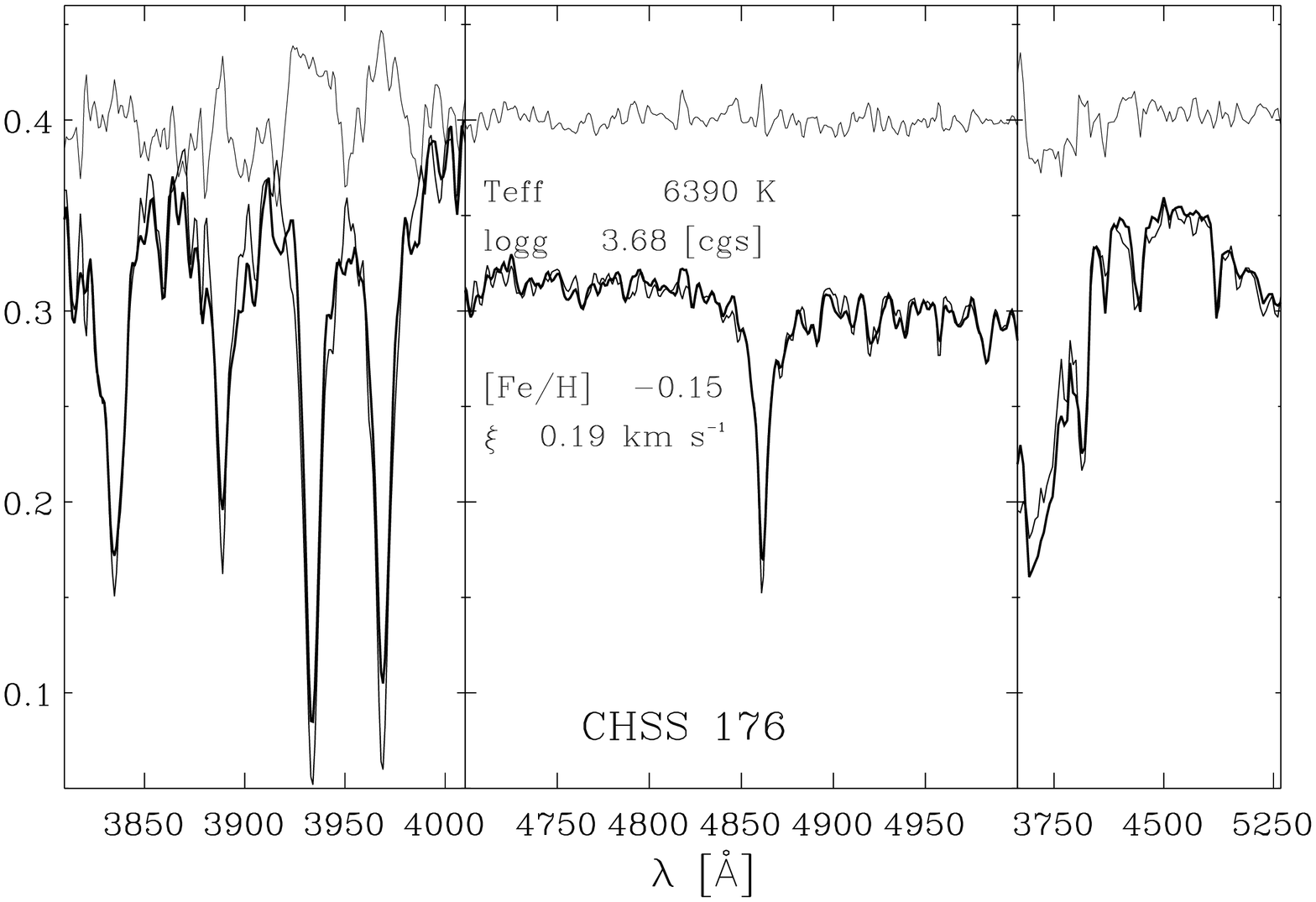}
    \includegraphics[width=2.5in, angle=90, clip=true]{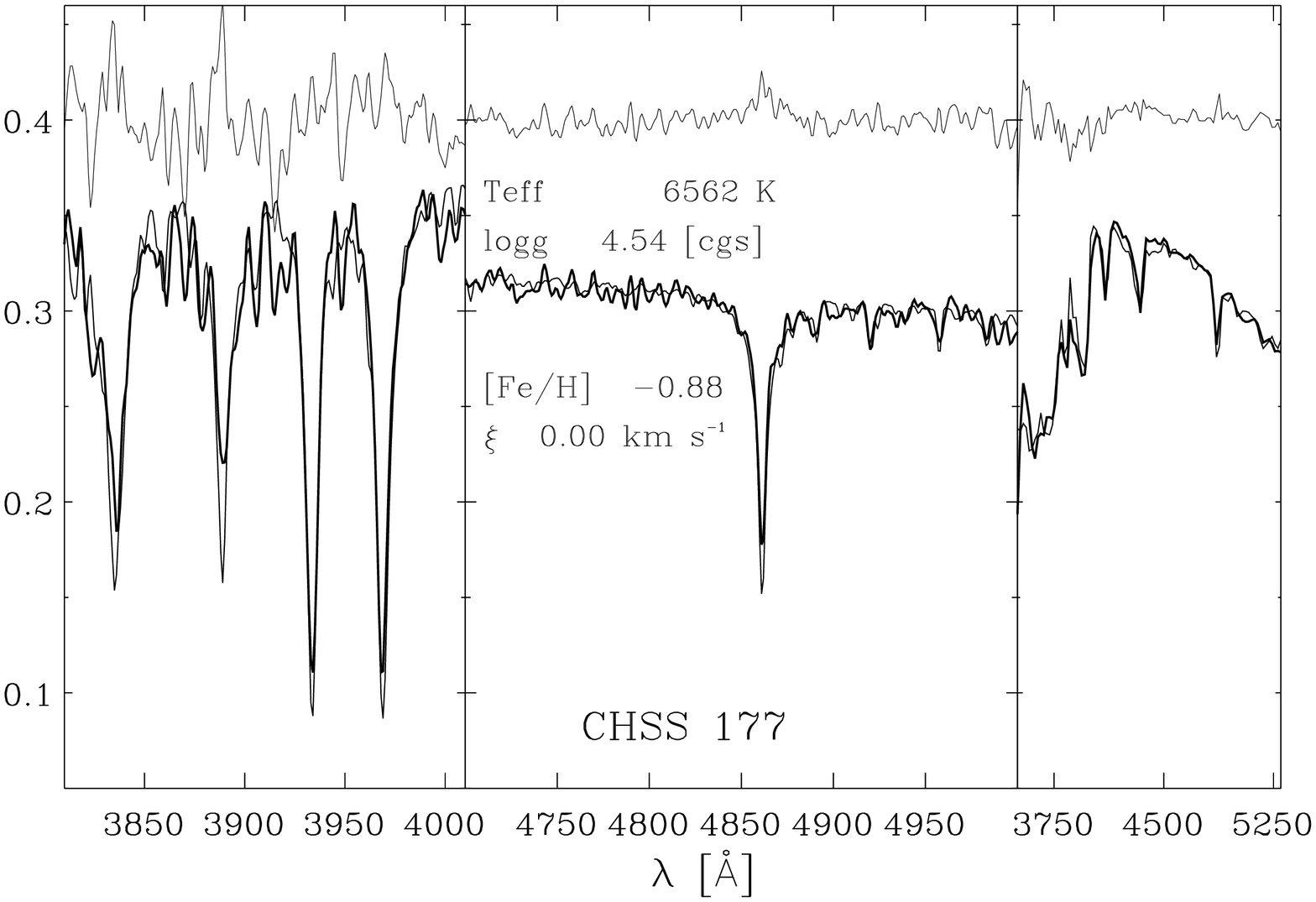}
 \figcaption{ \label{fitting}
	Comparison between the observed (thick line) and model (thin line)
spectra for four stars in the sample. We show the observed$-$model
difference (thinnest line) shifted by +0.4 above the spectra. The fluxes
have arbitrary units.  The different windows were independently
normalized.}

	~ \\

The star CHSS~118 is a good example of a warm metal-poor object. The
spectral lines of metallic species in the region around H$\beta$ are
almost totally lost in the noise; the strength of the Ca {\small II} K
line constrains the metallicity.  The strength of the Balmer lines is
irreconcilable with the continuum slope for this star, probably reflecting
a large systematic error in the spectrophotometric calibration.  The
algorithm gives a higher weight to the Balmer lines than the continuum
slope when determining effective temperature.  Note that our procedure
relies on {\it relative}, not absolute, fluxes, and we refer to these
relative fluxes as {\it spectrophotometry}.  We only rely on the absolute
flux calibration of the observed spectra in Section \ref{sec:dist} to
estimate distances.

\subsection{Estimation of (\bv)$_0$}

The (\bv)$_0$ color provides an estimate of effective temperature, and
allows us to access independent methods of measuring $T_{\rm eff}$,
$\log{g}$, and [Fe/H]. Although we lack measured $B$ photometry, we can
approximate (\bv)$_0$ from the combination of Balmer line strengths and
2MASS $(J-K)_0$ colors, where available. We label this color estimate
$BV0$ to distinguish it from an observed (\bv)$_0$.

First, we employ a neural network approach to estimate $BV0$ based on
observed stellar Balmer-line strengths. We take a large set of stars with
available (\bv)$_0$ colors and $HP2$ and $KP$ spectral indices from the HK
survey of Beers and collaborators to train the neural network. We include
only those stars with inferred reddening $E(\bv) \le 0.03$ in this
training set. Based on comparisons with an extensive validation set of
stars (not seen by the neural net during training and testing), we obtain
external 0.03-0.035 mag errors in the estimate of $BV0$, i.e., at the
level of the accuracy in the derived reddening corrections.

We use a second neural network to estimate $BV0$ from 2MASS colors. We
train the neural network using a large set of stars from the HK survey
with available $(J-K)_0$ and (\bv)$_0$ colors. Here we include only those
stars with $E(\bv) \le 0.03$ and with errors in the $J$ and $K$ magnitudes
$\le0.04$ mag. We train this simple neural network with $(J-K)_0$ as input
and (\bv)$_0$ as output, and obtain external errors in the predicted color
$BV0$ of $\sim$0.07 mag.

To obtain a final estimated $BV0$ color, we must first identify stars
likely to be bluer or redder than $(\bv)_0 = 0.0$, where the strength of
the Balmer lines are greatest, and for which a degeneracy exists blueward
and redward of this location when estimating colors from their strengths.  
We use He {\small I} lines to make this separation; these lines are
present in the bluer stars and absent in the redder stars. We obtain a
final weighted estimate from:  $BV0 = (3 BV0_{\rm HP2} + BV0_{\rm
2MASS}$)$/4$, where $BV0_{\rm HP2}$ is the prediction based on the
Balmer-line strengths and $BV0_{\rm 2MASS}$ is the prediction based on the
2MASS colors.  When 2MASS photometry is not available, we use the
$BV0_{\rm HP2}$ prediction as the final estimate of $BV0$.  Comparison
with the observed (\bv)$_0$ colors of the calibrator stars indicates that
the final external error in the estimated $BV0$ color is $\sim$0.04 mag
over the color range of the Century Survey stars reported in this paper,
and in any case certainly adequate for abundance determination.

\subsection{Comparison of Stellar Parameters and Errors}

Here we compare our values of $T_{\rm eff}$, $\log{g}$, and [Fe/H] with
the those derived from the independent methods of \citet{beers99} and
\citet{wilhelm99a}.  Although these additional methods rely in part on the
same color estimates (as described above), they are complementary to one
another since they are calibrated separately, using different sets of
comparison stars.  Thus, a reasonable estimate of the uncertainties in
$T_{\rm eff}$, $\log{g}$, and [Fe/H] results from a comparison of the
values from all three methods.

\subsubsection{Comparison of Stellar Metallicity Estimates}

First, we measure line indices for prominent spectral features
\citep{beers99}.  We then use the line indices to obtain estimates
of [Fe/H] and $BV0$. As a further check, for the stars with sufficiently
high S/N spectra, we have made use of the ``auto-correlation function''
approach \citep{beers99} based on calculations kindly performed for us by
John Norris. The auto-correlation function technique is particularly
valuable for obtaining stellar metallicity estimates for the cooler, more
metal-rich stars in our sample. For such stars, the Ca {\small II} KP
index can suffer from saturation effects. From the infrared-flux-method
calibrations of \citet{alonso96} and \citet{alonso99} we then derive
$T_{\rm eff}$ based on these input estimates.

Second, we follow \citet[][RW]{wilhelm99a} and determine $T_{\rm eff}$ for
hot ($T_{\rm eff} > 7000$ K) stars using 2MASS $JHK$ photometry and our
Johnson $V$ photometry.  We determine $\log{g}$ using a combination of the
H$\delta$ Balmer-line width (measured at 20\% below the local continuum
level) and the slope of the Balmer discontinuity.  We compare this
combination with a grid of synthetic spectra computed using ATLAS9 models
and the spectral synthesis routine SPECTRUM \citep{gray94}. We compute
metallicity for the hot stars by comparing the observations to a synthetic
grid of equivalent widths for the Ca {\small II} K line.  We also
performed a chi-square comparison between metallic-line regions in
synthetic and observed spectra \citep{wilhelm99a}.  This latter approach
provides valuable information required to identify (in particular)
metallic-line A-type stars with peculiar Ca {\small II} K line strengths.

Appendix Data Table \ref{tab:dat3} summarizes the individual metallicity
estimates from all three methods. Column (4) is the GA-derived
metallicity, denoted as [Fe/H]$_{\rm GA}$. Column (5) is the
\citet{beers99} derived metallicity based on the KP index, [Fe/H]$_{\rm
KP}$. Column (6) is the \citet{wilhelm99a} derived metallicity,
[Fe/H]$_{\rm EC}$.  The designation ``EC'' arises from the use of both
equivalent widths of spectral features and chi-square spectral matches in
deriving the metallicity by this method. Column (7) lists our best
estimate of stellar metallicity, [Fe/H]$_{\rm final}$.

Figure \ref{fig:fehcomp} shows that the metallicities of all three methods
are in good agreement. We find [Fe/H]$_{\rm KP} -$[Fe/H]$_{\rm GA} = 0.03
\pm 0.31$, comparing 550 cool stars with $BV0 >0.3$ mag (and accepted
abundance estimates by both techniques). Similarly, we find [Fe/H]$_{\rm
EC} -$[Fe/H]$_{\rm GA} = -0.09 \pm 0.37$, comparing 117 accepted
abundances for hot stars with $BV0 \le 0.3$ mag. Finally, we find
[Fe/H]$_{\rm EC} -$[Fe/H]$_{\rm KP} = -0.12 \pm 0.27$, comparing 536 cool
stars with $BV0 >0.3$ mag. We note that if we ignore the outliers (marked
with ``:''), the dispersions between the three methods reduce to $\pm0.21$
dex. Because the final metallicity is an average of two or three of the
methods, we believe [Fe/H]$_{\rm final}$ is accurate to $\pm0.25$ dex.

	~

 \includegraphics[width=2.9in]{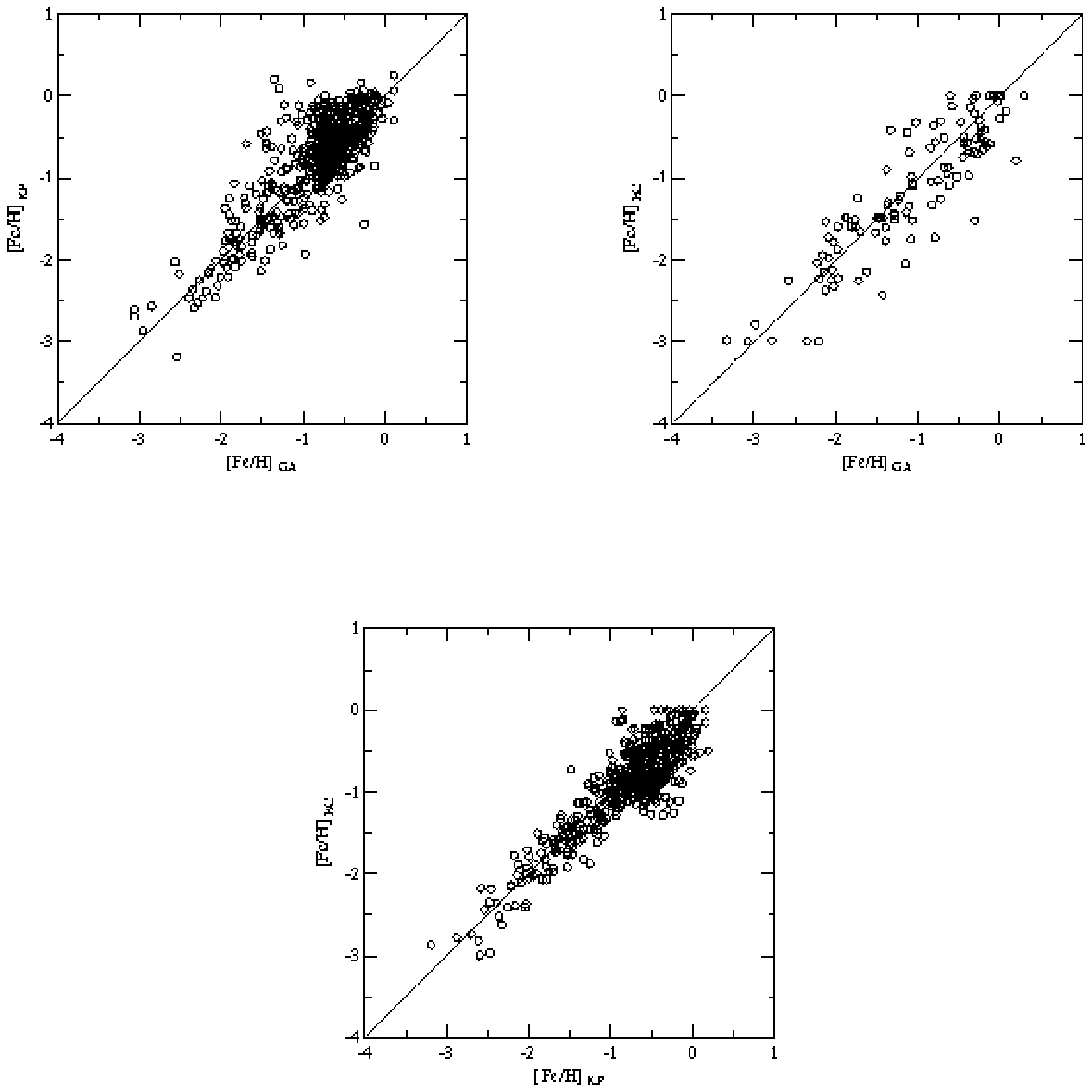}
	\figcaption{ \label{fig:fehcomp} Abundance comparison between the
three methods used to derive metallicity estimates, which we designate
[Fe/H]$_{\rm KP}$, [Fe/H]$_{\rm GA}$, and [Fe/H]$_{\rm EC}$. We conclude
that the abundance estimates that we employ are accurate to 0.25 dex (see
text).}

We employ all three methods to obtain [Fe/H]$_{\rm final}$. For cool stars
with $BV0 \ge 0.3$, the KP and GA methods are the primary metallicity
indicators. For the hot stars with $BV0 < 0.3$, the GA and EC methods are
the primary indicators. In most cases, we obtain [Fe/H]$_{\rm final}$ by a
straight average of the appropriate primary indicators, chosen from the
three methods (when available). Occasionally (due to either a low S/N
spectrum, poor flux calibration, or absent photometry), the primary
metallicity estimates are strongly discrepant with one another. We then
use the third available method as a tie breaker, and average it with the
other indicator with which it best agrees. For a small number of the hot
stars where the KP method cannot be used, there remains a large
discrepancy between the GA and EC estimates; hence we made choices based
on our best guess of which metallicity indicator was led astray (poorly
fit spectra, for example, would be one reason to reject the GA estimate).
When the final average involves metallicities that disagree by more than
0.4 dex, or when we use only one of the three methods, we indicate some
additional uncertainty with a ``:'' next to [Fe/H]$_{\rm final}$. There
are a small number of stars where, even after carrying out these
procedures, visual inspection of the stellar spectra suggested that the
derived final estimate of metallicity was suspect (for example,
identification as a metallic-line A star). In such cases we replaced the
final metallicity estimate with one that we feel is more likely to be
correct than the individual methods suggested (e.g., when we thought it
likely that the star was of solar metallicity).

\subsubsection{Comparison of $T_{\rm eff}$ Estimates}

The temperature estimates obtained from the spectrophotometric and
{\it photometric} $T_{\rm eff}$s differ significantly:  the GA-derived
spectrophotometric $T_{\rm eff}$ is higher by 257 K ($\sigma=196$ K).  
Because the photometric scale is more robust, we decrease the
spectrophotometric $T_{\rm eff}$ by 257 K.  We can obtain a third estimate
of the effective temperature by combining our $V$ magnitudes with the $K$
magnitudes measured by 2MASS.  We use the \citet{alonso99} calibrations
for this color. We obtain a mean difference $T_{\rm eff} (BV0) - T_{\rm
eff} (V-K) = 116 \pm 9$ K ($\sigma = 212$ K).  A bias results from the
different zero point between the 2MASS and Johnson $K$ bandpasses.  We
thus correct the $(V-K)$-based $T_{\rm eff}$s to the $BV0$ scale. We
note that the \citet{alonso99} scales are limited to $5000 \lesssim T_{\rm
eff} \lesssim 8000$ K; we cannot check for systematic differences outside
this range.

Figure \ref{phot} compares the photometric and spectrophotometric
temperatures. The solid lines indicate the mean shifts.  Whenever the
$BV0$ or $(V-K)$ calibrations could be used, we considered them in the
final $T_{\rm eff}$ estimate.

\subsubsection{Comparison of $\log g$ Estimates}

We finally compare the GA-derived gravities with those determined with the
\citet{wilhelm99a} method described here.  The GA gravities are higher by
0.26 dex, with a standard deviation of 0.35 dex.  We deem this difference
reasonably good, considering the difficulty of estimating surface gravity
from medium-resolution spectra.  We adopt the GA-derived gravities as
our final value of $\log g$ and estimate the error to be 0.25 dex.

In summary, our values of $T_{\rm eff}$, $\log g$, and [Fe/H] are
uncertain by roughly 200 K, 0.4 dex, and 0.25 dex, respectively. Appendix
Data Table \ref{tab:dat3} lists the adopted values of the physical
parameters for our program stars.

\section{SPECTRAL CLASSIFICATION} \label{sec:class}

In addition to the stellar parameters described in \S \ref{sec:pars},
stellar spectral classification provides a useful way to organize the
stars in our sample.  BHB stars, for example, typically have an early
A-type spectral classification, while the cooler stars are typically of
later classes.  We perform spectral classification by
measuring the relative strengths of absorption lines in the stellar
spectra, and make use of the stellar spectral line indices of
\citet{oconnell73} and \citet{worthey94}.  We determine spectral types
from line-index versus spectral-type relations that we have derived for
the \citet{jacoby84} library of (luminosity class V) stellar spectra.

Table \ref{tab:lines} summarizes the line-index versus spectral-type
relations we employ. Our spectral classifications are based primarily on
the Ca {\small II} 3933 \AA\ line index and the H$_{\rm sum}$ line
index---a sum of the H 3889 \AA, H 4101 \AA, H 4340 \AA, and H 4861 \AA\
line indices.  As an example, Figure \ref{fig:caiiline} shows the Ca
{\small II} line-index bands \citep{oconnell73} overlaid on an A and an F
stellar spectrum from our dataset. We use the CN 3860 \AA, CH 4305 \AA, Mg
{\small I} 5175 \AA, and Fe$_{\rm sum}$ line indices to extend our
spectral classification range and to corroborate the type obtained from
the Ca {\small II} and H$_{\rm sum}$ line indices. In practice, each
spectral classification typically depends on four spectral line indices,
and has an uncertainty of $\pm1.1$ spectral sub-types.

 \includegraphics[width=3.25in, angle=0, clip=true]{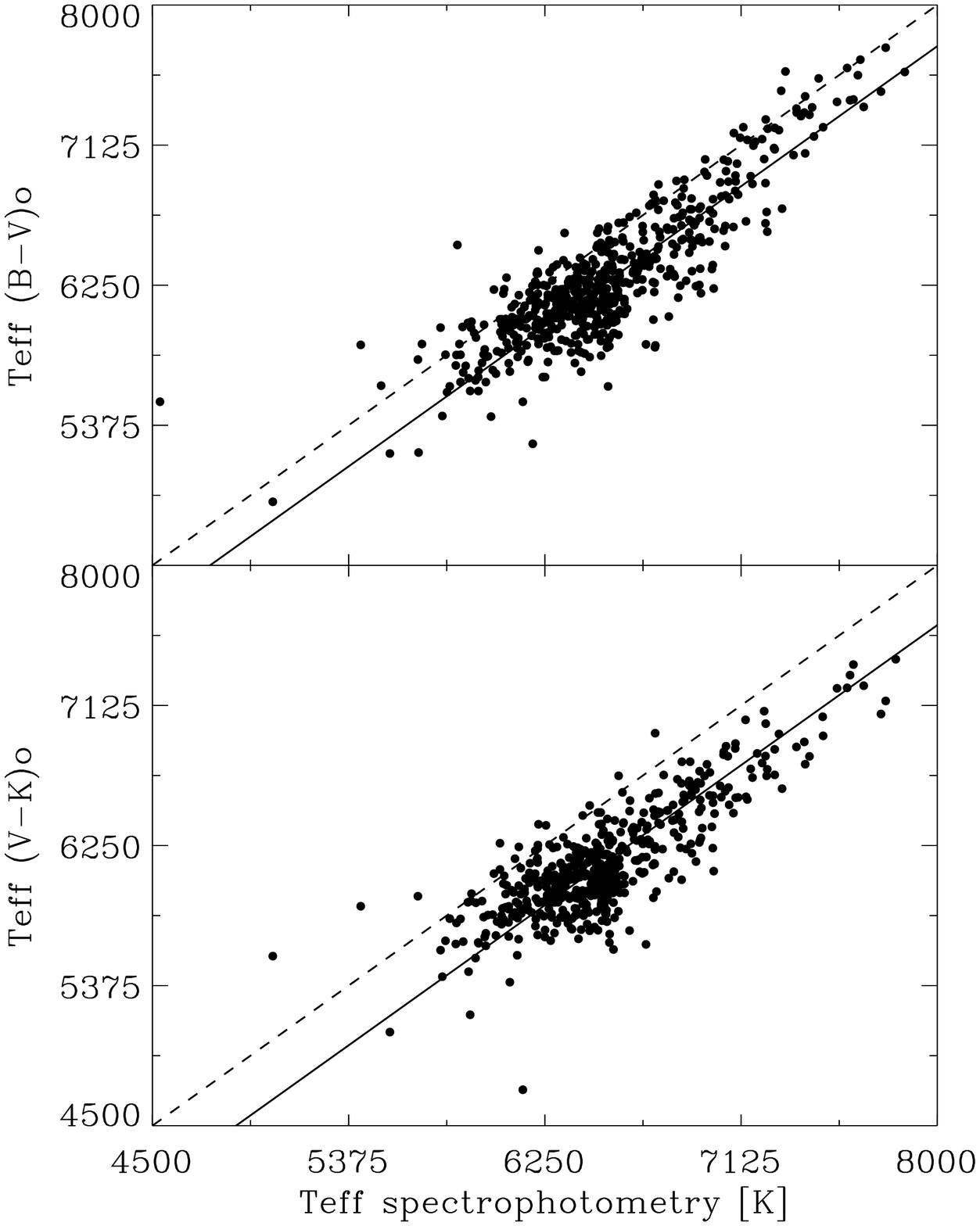}
 \figcaption{ \label{phot}
	The photometric $T_{\rm eff}$s plotted against the GA-derived
spectrophotometric $T_{\rm eff}$.  The dashed lines have a slope of one.
The solid lines indicate the mean shifts.}

 \includegraphics[width=3.25in, bb=24 165 590 690]{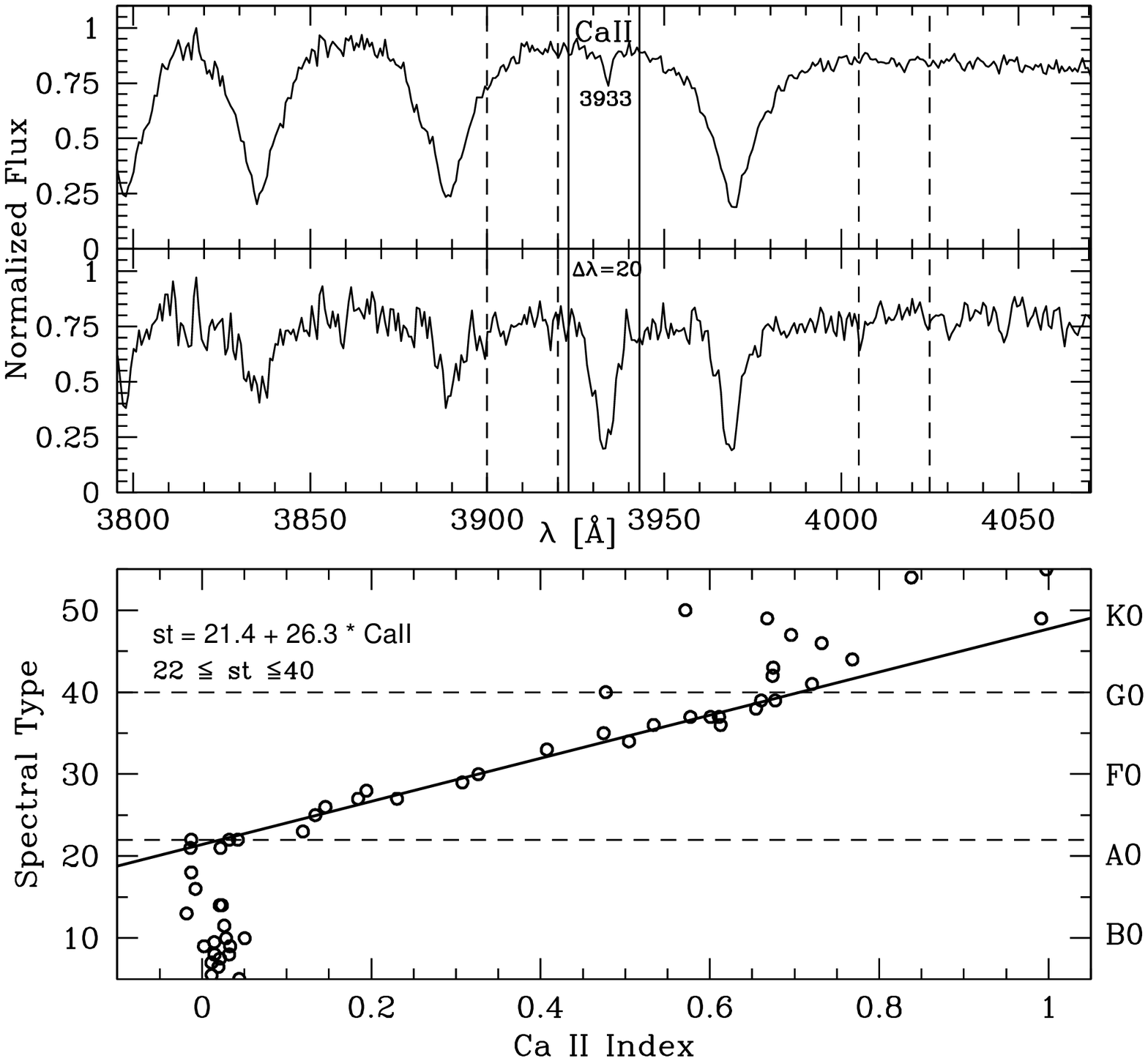}
 \figcaption{ \label{fig:caiiline}
	Upper panel:  the \citet{oconnell73} Ca {\small II} 3933 \AA\
spectral line-index band and sidebands, plotted on an A and F star
spectra.  Lower panel:  the line index values measured for the
\citet{jacoby84} library of luminosity class V stellar spectra.  The 
spectral type relation is valid between A2 and G0, with a residual of
$\pm1.1$ spectral sub-types.}

	~

Because the \citet{jacoby84} library largely contains solar-metallicity
stars, the line index versus spectral type relations are appropriate for
stars of solar metallicity.  There are systematic offsets in spectral
types for the metal poor stars; as a result the effective temperatures for
the most metal-poor stars in our sample are 1000 K too low for the
spectral types we measure.  Recognizing this difficulty with the spectral
types, we use them only to separate A-type stars from F-type stars, and to
identify unusual objects.

\section{DISTANCE ESTIMATES} \label{sec:dist}

We estimate distances to our stars in two ways. For the likely BHB stars
in the sample, we employ our broadband photometry and published
luminosity-metallicity relations to provide reasonably precise distance
estimates. For the rest of the sample, we use estimates of stellar angular
diameter and stellar radius from theoretical isochrones to provide rough
distance estimates.

\subsection{BHB Distances}

Published $M_V$-metallicity relations for horizontal branch stars vary
considerably, with values of the slope ranging between 0.15
\citep{carney92} and 0.30 \citep{sandage93}.  Values of the zero point
fall into two groups, $\sim$0.30 mag apart.  Fortunately, a search for
star streams only requires {\it relative} distances.  Thus we do not concern
ourselves with the exact zero-point of the $M_V$-metallicity relation.

We use the {\it Hipparcos}-derived zero point, $M_V(RR)=0.77\pm0.13$ mag
at [Fe/H] = $-$1.60 \citep{gould98}, based on the statistical parallax of
147 halo RR Lyrae field stars. We employ the recently measured
$M_V$-metallicity slope $0.214\pm0.047$ \citep{clementini03}, based on
photometry and spectroscopy of 108 RR Lyrae stars in the Large Magellanic
Cloud. We use the resulting relation, $M_V(RR) = 0.214 {\rm [Fe/H]} +
1.11$ mag, to calculate distances to the BHB stars. We note that this
empirical relation yields absolute magnitudes $0.21\pm0.03$ mag less
luminous than the theoretical ZAHB relations of \citet{caloi97},
\citet{salaris97}, and \citet{cassisi99} over the metallicity range $-3 <
{\rm [Fe/H]} < 0$ of our BHB stars. Appendix Data Table \ref{tab:dat3}
lists the distances for the BHB stars.


\subsection{Non-BHB Distances}

We derive rough distances to the non-BHB stars in our sample using
estimates of stellar angular diameter, $\theta$, and stellar radius, $R$.  
The absolute fluxes and the derived stellar parameters for our stars allow
a direct estimate of the stellar angular diameter from:
	\begin{equation}
 \theta \simeq
 2  \sqrt{\frac{f_{\lambda}}{F_{\lambda}(T_{\rm eff}, \log g, {\rm [Fe/H]})}}
\end{equation}
	\noindent where $f_{\lambda}$ is the observed flux at Earth and
$F_{\lambda}(T_{\rm eff}, \log g, {\rm [Fe/H]})$ is the flux at the
stellar surface predicted by a model atmosphere. We can obtain a second
estimate of the angular diameter $\theta$ by following
\citet{dibenedetto98}, who derived a Barnes-Evans relation between the
surface brightness and the $(V-K)_0$ color for spectral types in the range
$\sim$ A0 to G2. Before applying this approach, we correct for the
systematic offset of 0.07 mag in the $(V-K)_0$ color as explained in \S
\ref{sec:pars}.

Stars at a given locus in the HR diagram have a limited range of stellar
radius $R$. By using stars in eclipsing binary systems with radii known to
$<2\%$, \citet{allende99} find that the position in the (\bv)--$M_V$ plane
can constrain the stellar radius $R$ to $\sim$8\% for nearby
solar-metallicity stars. Similarly, the position of a star in $T_{\rm
eff}$-$\log{g}$-[Fe/H] space provides a useful means to estimate the
stellar radius $R$, and we exploit our spectroscopic determinations of
these parameters to this end. We use the isochrones of \citet{bertelli94}
to relate $T_{\rm eff}$, $\log{g}$, and [Fe/H] to $R$. We adopt normal
error distributions for $T_{\rm eff}$ and $\log{g}$, and compute a
weighted average of $R$ values that fall within 3-$\sigma$ of a star's
($T_{\rm eff}$, $\log{g}$).  We only use isochrones within 0.4 dex of a
star's [Fe/H] for this calculation.  \citet{reddy03} describe in more
detail the same technique applied to derive stellar ages.

We calculate the final distance by combining the radius, $R$, with the two
estimates of angular diameter, $\theta$.  Appendix Data Table
\ref{tab:dat3} lists the distances derived from the \citet{bertelli94}
isochrones for the non-BHB stars.  Distances based on the
\citet{dibenedetto98} $S_V$-$(V-K)_0$ relationship are on average 20\%
smaller than those from the \citet{bertelli94} isochrones, with an RMS
scatter of 25\%.  This internal uncertainty is only a lower limit to the
errors in the derived distances, because the two methods share an
important error source, the radius estimate.  The flux calibration of the
spectra introduces additional uncertainty, so these distances should be
treated as rough estimates only.

%

\section{BHB CLASSIFICATION} \label{sec:bhb}

	BHB stars comprise only a subset of our blue star sample. Thus we
require a method to estimate surface gravity that is sufficiently accurate
to separate the BHB stars from the A dwarfs and blue stragglers also
present in our sample.


	We use four recent techniques to refine our selection of BHB
stars. \citet{kinman94} develop the $\Lambda$ classification, which
combines spectrophotometric measures of the size and steepness of the
Balmer jump. \citet{wilhelm99a} use photometric and spectroscopic measures
of $T_{\rm eff}$, $\log{g}$, and [Fe/H] to select BHB stars. Most
recently, \citet{clewley02} use a Balmer-line-width-color technique and a
Balmer line-shape technique that reliably reproduce the \citet{kinman94}
$\Lambda$ classification without the need for spectrophotometry. Below, we
estimate errors in the application of these techniques to the blue program
stars. We compare the results of all four techniques, and make our final
selection of BHB stars based on this comparison.

\subsection{Kinman et al.\ Method}

We begin by investigating measures of the size and steepness of the Balmer
jump to discriminate BHB stars. \citet{kinman94} define a
spectrophotometric index, BJA$_0$, that quantifies the size of the Balmer
jump, as well as a parameter, $\lambda_{0.5}$, that measures the slope of
the Balmer jump. Both parameters are sensitive to surface gravity. In
Table \ref{tab:kinman} we compare the values of BJA$_0$ and
$\lambda_{0.5}$ that we obtain for 33 BHB/A stars in common with their
sample. Our BJA$_0$ values agree to within a scatter of 5\%. However, our
values of $\lambda_{0.5}$ are discrepant by $30 \pm 40\%$. The large
offset and scatter in $\lambda_{0.5}$ is likely caused by differential
atmospheric refraction (we did not observe at the parallactic angle); thus
$\lambda_{0.5}$ is of little use for our application.

The Balmer lines provide another discriminant of BHB stars. A combination
of BJA$_0$ and our spectral type, which is largely driven by the H$_{\rm
sum}$ line index for A stars, discriminates well between the 33
\citet{kinman94} BHB and A stars that we observed. In our sample there are
41 BHB candidates with BJA$_0 >1$ mag and spectral type earlier than A2.

\subsection{Wilhelm et al.\ Method}

Next, we investigate using effective temperatures and surface gravities to
discriminate BHB stars, following the methods of \citet{wilhelm99a}.

BHB stars have a well-determined location in the $\log{g}$ - $T_{\rm eff}$
plane.  The maximal $\log{g}$ for a ZAHB star is $\log{g} < (4.33
\log{T_{\rm eff}} - 13.23)$ \citep{sweigart87}. We assign a BHB
classification to stars that meet this criterion. Note, however, that the
surface gravities of main-sequence stars overlap the surface gravities of
BHB stars at large $T_{\rm eff}$. To understand where the $\log{g}$'s
overlap, we select ``normal'' main-sequence A-stars (stars that do not
have variability or peculiar metal abundances) from the catalog of
\citet{cayrel97} and find that the minimal $\log{g}$ for main-sequence A
stars follows the bounding line $\log{g} = 4.79 - 1.11\times10^{-4} T_{\rm
eff}$. This relation gives $\log{g} = 3.7$ for $T_{\rm eff} = 9900$ K.
Thus a star with $\log{g} > 3.7$ is classified as a main-sequence star
unless it overlaps at the hot end with the BHB trend line, in which case
it is labeled BHB/A. We identify 67 BHB candidates with the
\citet{wilhelm99a} technique.

\subsection{Clewley et al.\ Methods}

Finally, we investigate using the ``$D_{0.15}$-Color'' and ``Scale
width-Shape'' methods of \citet{clewley02} to discriminate BHB stars from
other A-type stars. These methods rely on fitting a Sersic profile to the
H$\gamma$ and H$\delta$ Balmer lines. The Sersic profile $y=1-a \exp{[(|
x-x_o|/b)^c]}$ depends on 3 parameters, the line depth, $a$, the scale
length, $b$, and the power, $c$, of the profile. We closely follow the
procedure described by \citet{clewley02} for fitting the H$\gamma$ and
H$\delta$ profiles, masking metallic blends and iteratively rejecting
outliers. Like \citet{clewley02}, we find that a convolution of a Sersic
function and a Gaussian with the FWHM of the Th and Ar lines (2.30 \AA)
reproduces the observed line shapes very well.

The $D_{0.15}$-Color method separates BHB stars based on the width of the
Balmer lines (measured at 0.85 of the continuum flux level) and the
$(\bv)_0$ color. The line width is sensitive to surface gravity, while the
color is (primarily) sensitive to effective temperature.  We compute the
value of $D_{0.15}$ from the Sersic profile fit, $D_{0.15}=2b
\ln{(a/0.15)}^{1/c}$, where the line depth $a=0.83$. We find 56 BHB
candidates with the $D_{0.15}$-Color method.

The Scale width-Shape method separates BHB stars based on the scale length
$b$ of the Sersic profile, which provides a measure of surface gravity,
and the power, $c$, which provides a measure of effective temperature. BHB
stars have $b<25.7c^3 -138.2c^2 + 187.5c - 66$ (Clewley 2002, private
communication), where $b$ is in units of \AA\ and $c$ in units of the
normalized flux.  We find 73 candidates with the Scale width-Shape method.

According to \citet{clewley02}, the $D_{0.15}$-Color method should select
BHB stars with a completeness of $\sim$87\% and a contamination rate of
$\sim$7\% for spectra with S/N=15. At the same S/N, the Scale-width Shape
method should select BHB stars with a completeness of $\sim$82\% and a
contamination rate of $\sim$12\%. This efficiency is similar to the
claimed efficiency of BHB selection by \citet{wilhelm99a}.

 \includegraphics[width=3.35in,bb=19 155 575 696,clip=yes]{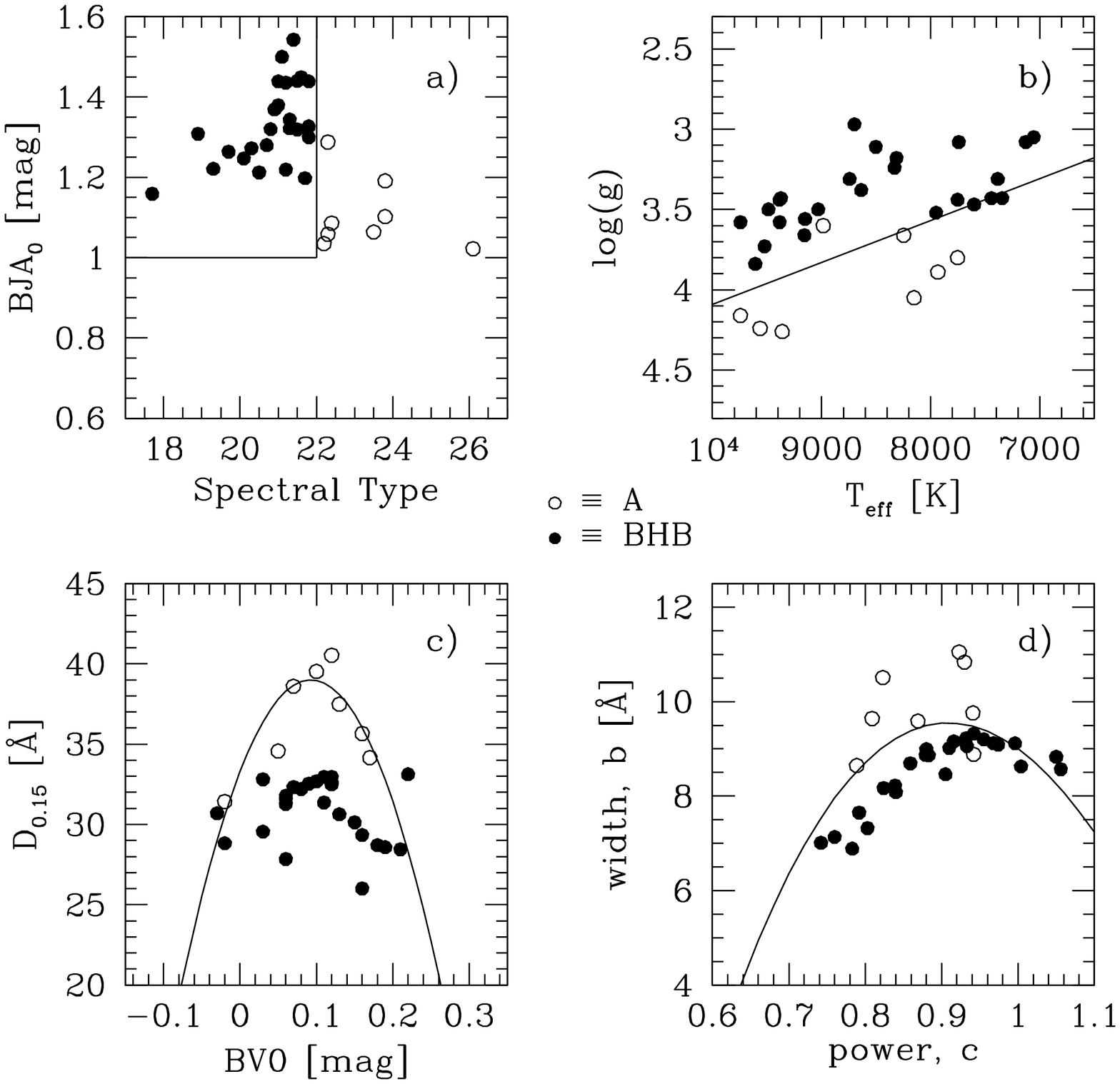}
 \figcaption{ \label{fig:bhba}
The four BHB classification methods applied to the 33 BHB/A stars
observed in common with \citet{kinman94}:  a) the modified
\citet{kinman94} method, b) the \citet{wilhelm99a} method, c) the
\citet{clewley02} $D_{0.15}$-Color method, and d) the \citet{clewley02}
Scale width-Shape method.  Solid circles mark the BHB stars; open circles
mark the A stars.}

 \includegraphics[width=3.35in,bb=19 155 575 696,clip=yes]{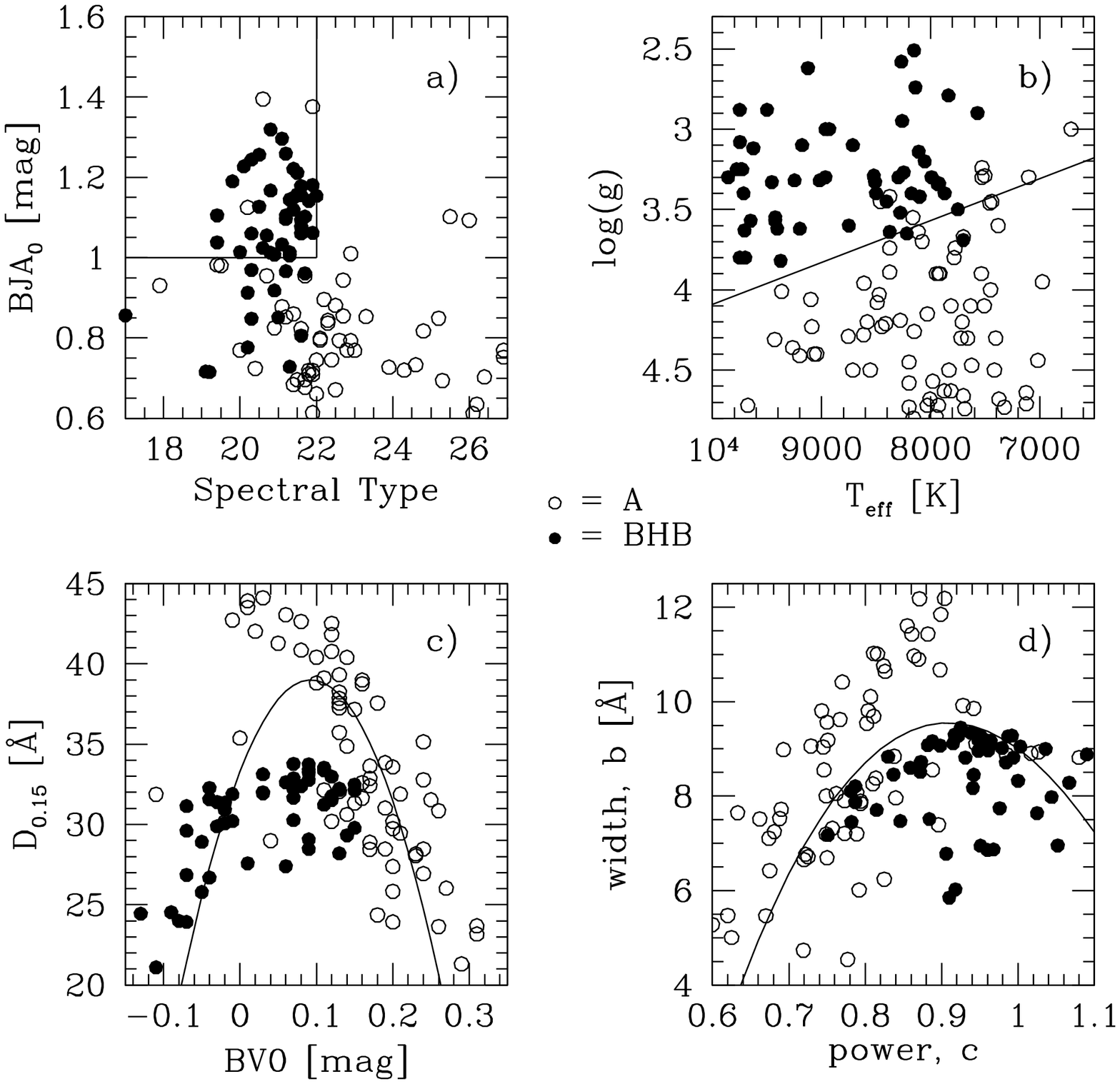}
 \figcaption{ \label{fig:bhbaa}
The four BHB classification methods applied to the 130 early
A-type stars in our sample.  The panels are the same as in Figure
\ref{fig:bhba}.  Solid circles mark the BHB stars, stars positively
classified by three or more of the four methods; open circles mark the
other A stars.}

\subsection{Final BHB Sample}

All four methods we employ for the identification of BHB stars are
sensitive to uncertainties in the parameters derived for the stars. To
obtain the purest BHB sample possible, we examine the output samples from
each method. Figures \ref{fig:bhba} and \ref{fig:bhbaa} show the results
of the four methods for the 33 BHB/A stars in common with \citet{kinman94}
and for 130 early A-type stars from the Century Survey Galactic Halo
Project. The \citet{wilhelm99a} method in Figure \ref{fig:bhbaa}b appears
to provide the cleanest classification. The \citet{clewley02} methods in
Figure \ref{fig:bhbaa}c and \ref{fig:bhbaa}d appear to degrade at lower
effective temperatures (redder $BV0$ and lower power $c$).

There are 18 BHB stars in common to all four methods. There are an
additional 37 stars selected by three of the four methods. Thus a total
sample of 55 BHB stars is positively classified by three or more of the
four methods.

Because all four classification methods start to fail at high
temperatures, we can reduce contamination from blue stragglers (at the
expense of selecting fewer BHB stars) by imposing a $BV0>0$ mag color
limit \citep[see][]{kinman94, clewley02}. The BHB sample contains 33 stars
in this case. However, the hot $BV0<0$ BHB candidates all have low
$\log{g}$ and large BJA$_0$ values. We thus choose not to impose an
a-priori color limit.

Table \ref{tab:bhbaa} summarizes the BHB selection parameters for the 130
early A-type stars in the Century Survey Galactic Halo Project. Column (1)
lists the star identification. Column (2) lists our spectral type and
column (3) lists the BJA$_0$ value. Column (4) lists a ``1'' if the star
is classified as BHB by the modified \citet{kinman94} method (Figure
\ref{fig:bhbaa}a), otherwise it is zero. Column (5) lists the effective
temperature and column (6) lists the surface gravity. Column (7) lists a
``1'' if the star is classified as BHB by the \citet{wilhelm99a} method
(Figure \ref{fig:bhbaa}b), otherwise it is zero. Column (8) lists the
$BV0$ color and column (9) lists the $D_{0.15}$ value. Column (10) lists a
``1'' if the star is classified as BHB by the \citet{clewley02}
$D_{0.15}$-Color method (Figure \ref{fig:bhbaa}c), otherwise it is zero.
Column (11) lists the power $c$ and column (12) lists the scale length $b$
of the Sersic profile fit. Column (13) lists a ``1'' if the star is
classified as BHB by the \citet{clewley02} Scale width-Shape method
(Figure \ref{fig:bhbaa}d), otherwise it is zero.  Column (14) is the sum
of the four BHB classification methods.  We consider the 55 stars selected
by three or more of the four methods as BHB stars.

Table \ref{tab:bhba} summarizes the properties of the 33 BHB/A stars we
observed in common with \citet{kinman94}. The columns are identical with
Table \ref{tab:bhbaa}, with the addition of column (15), which lists the
\citet{kinman94} classification.  All BHB stars are correctly classified
as BHB by three or more of the methods, with the exception of HD064488.
\citet{kinman94} report that HD064488 was formerly classified as a BHB
star but is now known to have a high rotation.

\section{RESULTS} \label{sec:results}

Figures \ref{fig:plotsample} and \ref{fig:zfeh} show the physical
properties of the Century Survey Galactic Halo project stars.  Our sample
includes thin disk, thick disk, and halo stars.

Figure \ref{fig:plotsample} summarizes the spectral types and the radial
velocities (with respect to the local standard of rest) of our sample.
Sixty-four percent of the 764 blue stars are F-type stars and twenty-seven
percent are A-type stars.  This Figure shows a decreasing number of stars
from the F-types to the A-types, until the BHB stars (marked with filled
circles) appear around spectral type A1.  Twelve percent of our sample (92
stars) have spectral types between B9 and A3; 55 of these stars are BHB
stars.  The measured radial velocity dispersion of the BHB stars with
respect to the local standard of rest ($\sigma = 97$ km s$^{-1}$) is
consistent with a halo population.

	~ \\

 \includegraphics[width=3.5in, bb=20 153 563 690]{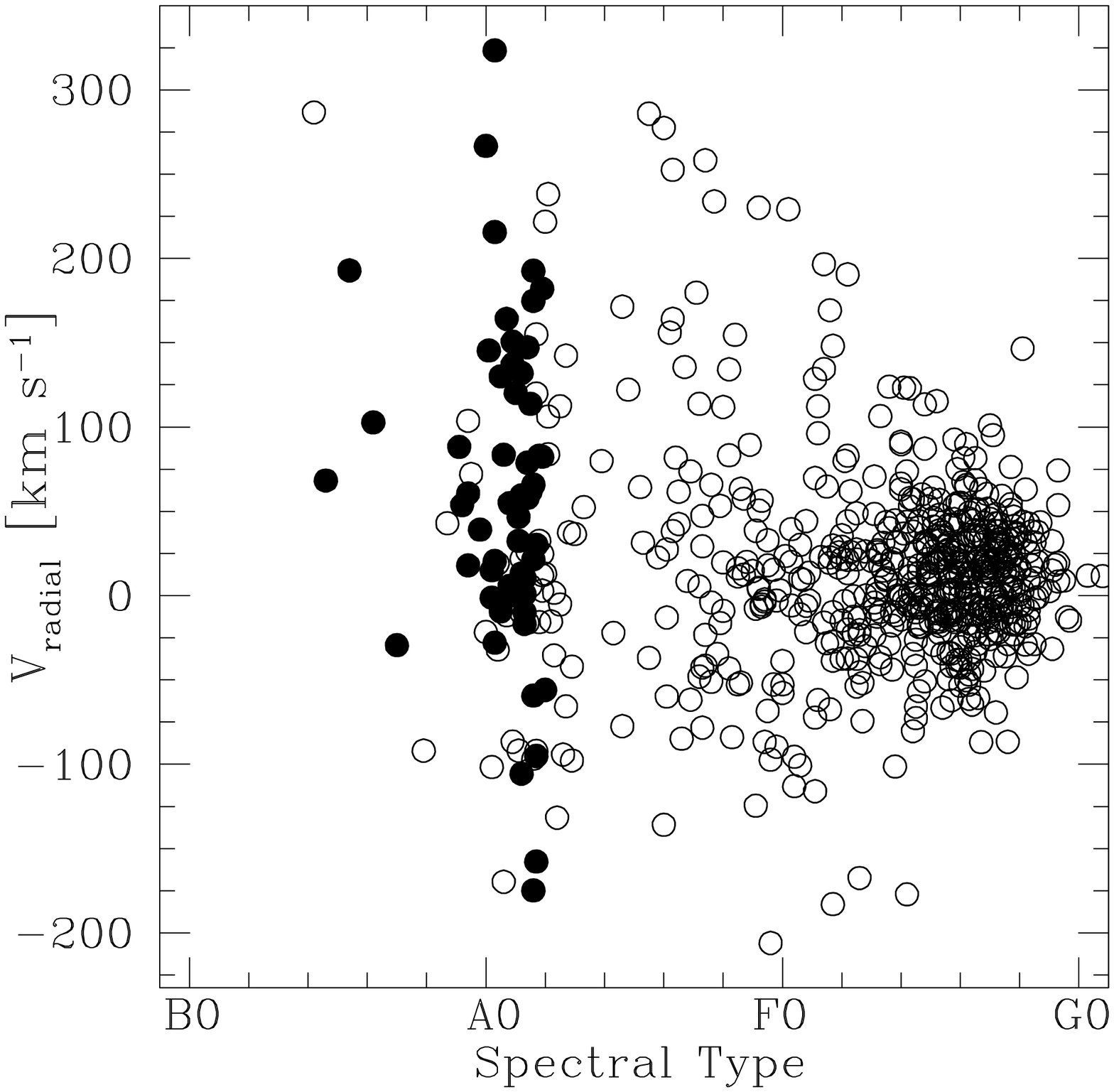}
 \figcaption{ \label{fig:plotsample} The spectral types and radial
velocities with respect to the local standard of rest in our sample.
Filled circles mark the BHB stars.}

 \includegraphics[width=3.6in, bb=59 155 578 690]{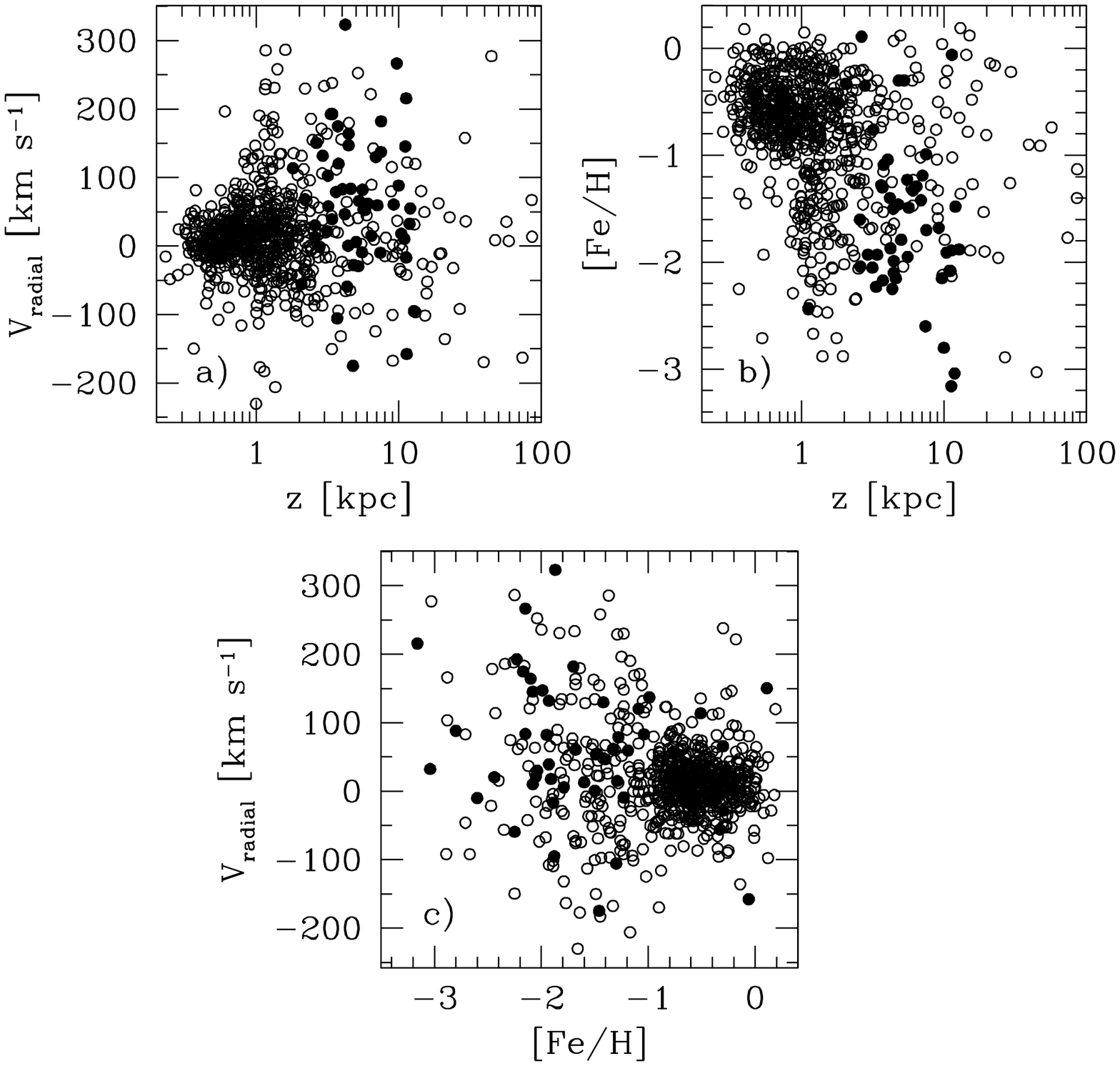}
 \figcaption{ \label{fig:zfeh} The distributions of radial velocity (with
respect to the local standard of rest), metallicity, and $z$ distance
above the Galactic plane in our sample.  Filled circles mark the BHB
stars.}

	~

Figure \ref{fig:zfeh} shows the velocities, metallicities, and distances
above the Galactic plane, $z$, for our sample.  The clump of stars with
$z<1$ kpc and [Fe/H]$>-1$ (Figure \ref{fig:zfeh}, upper right) are the
late F-types, with a mean $BV0=0.45$ mag, and radial velocity dispersion
with respect to the local standard of rest $\sigma = 31$ km s$^{-1}$.  
This velocity dispersion is consistent with the $\sim30$ km s$^{-1}$
dispersion measured for $(\bv)=0.45$ mag stars in the {\it Hipparcos} data
\citep{dehnen98}, suggesting that the late F-types are dwarf stars in the
thin and thick disks.  The stars with distances extending to $\sim$100 kpc
have surface gravities $\log{g}\sim2.0$ and are likely giants.  Our search
identifies BHB stars out to a distance of 13 kpc from the Sun.  The BHB
stars show a wide range of metallicity, $-3<$ [Fe/H] $<0$.


\subsection{Unusual Objects}

One of the benefits of examination of a large sample is the opportunity to
identify ``unusual'' objects.  There are 34 objects in our sample that do
not have typical A or F-type spectra.  We now examine these objects in
detail.

Six of the unusual objects are white dwarfs. \citet{mccook99} list five
of the white dwarfs as hot DA white dwarfs; one white dwarf appears to be
a new discovery.  Based on the \citet{wesemael93} white dwarf spectral
atlas, we classify the object at $13^{\rm h}10^{\rm m}13.4^{\rm s},
29^{\circ}43\arcmin59\arcsec$ (J2000) as a DZ7 white dwarf (Figure
\ref{fig:uspectra}).

Four of the unusual objects have normal B-type spectra.  \citet{green86}
identify three objects, for which we measure $\log{g}\sim2.4$, as
horizontal branch B stars.  Our BHB classification also selects these
three objects as BHB.  The fourth object has a dwarf-like $\log{g}=3.5$,
and a spectral classification of B8.5 (Figure \ref{fig:uspectra}). If this
star is a main-sequence star, its luminosity places it 0.8 kpc above the
Galactic plane.  \citet{conlon90} publish 32 B stars with similar
distances above the Galactic plane, in the range $0.5<z<4$ kpc.  A late B
star has $\sim$3.5 $M_{\sun}$ \citep{allen00} and a main-sequence lifetime
of $\sim2\times10^8$ years \citep{bowers84}.  Our B star has a radial
velocity +46 km s$^{-1}$.  It has traveled at least 0.8 kpc in
$2\times10^7$ years, consistent with its inferred lifetime if it formed
in the Galactic plane.

Eleven of the unusual objects are hot subdwarfs.  \citet{green86}
classify five subdwarfs as sdB and one as sdO.  We follow this
classification scheme, and classify the other five objects as sdA, based
on their high surface gravities, $\log{g}\sim5$, and the fact that their
Balmer-series lines merge into their continua (Figure \ref{fig:uspectra}).

Three of the unusual objects appear to be carbon-enhanced stars. These are
identified as stars lying well off the locus of GP vs.\ KP, shown as
filled circles in Figure \ref{fig:carbon}. In recent large-scale surveys
for metal-poor stars, such carbon-strong stars are found with increasing
frequency amongst the lowest metallicity stars \citep{norris97,rossi99}.
The estimated metallicities of the three carbon-enhanced stars in our
sample, are not, however, particularly low, falling in the range $-1.5 <
{\rm [Fe/H]}_{\rm F} \le -1.2$.

 \includegraphics[width=2.8in, bb=30 150 585 690]{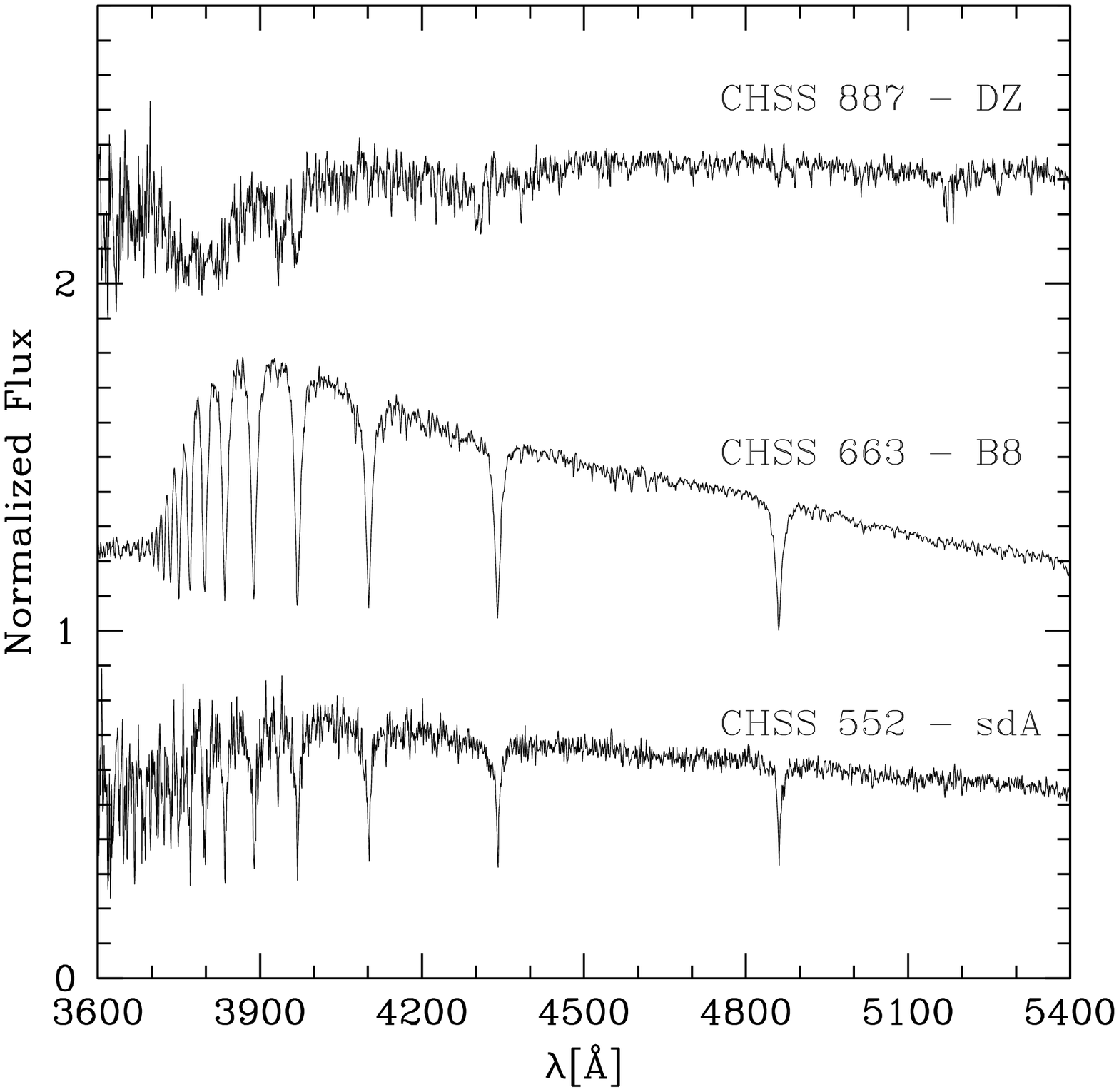}
 \figcaption{ \label{fig:uspectra} The spectra of the DZ7 white dwarf
CHSS~887, the B8.5 star CHSS~663, and the sdA object CHSS~552.}

 \includegraphics[width=2.6in]{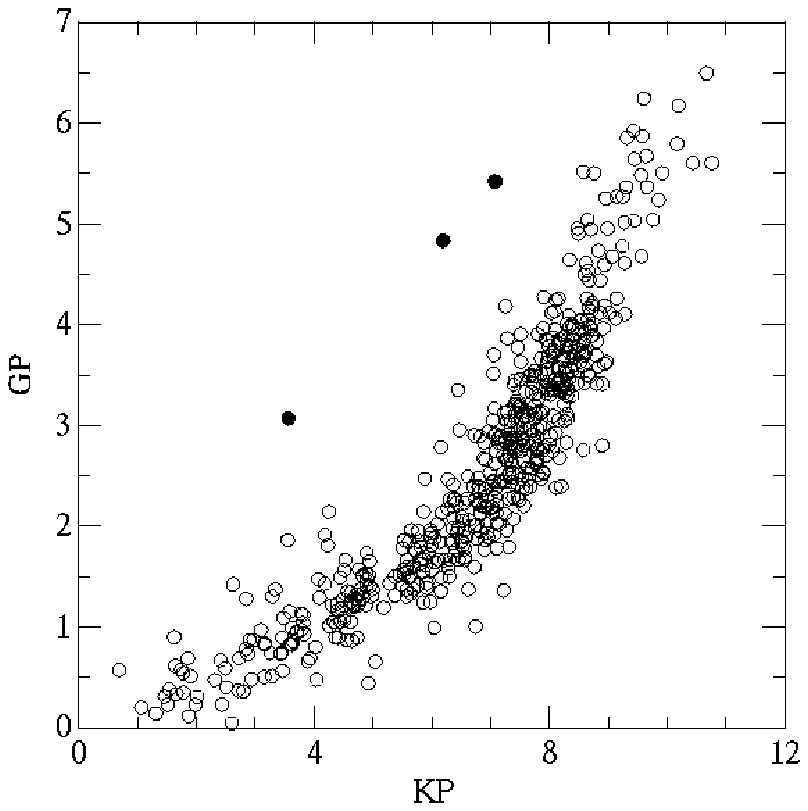}
 \figcaption{ \label{fig:carbon} GP, the G-band index of \citet{beers99},
as a function of KP, the CaII K index.  The three carbon-enhanced stars,
CHSS~197, CHSS~420, and CHSS~538, are indicated with filled circles.}

Of the ten remaining unusual objects, two appear to be composite spectra
with an F-type star plus a star with an earlier spectral type.  Six of the
unusual objects exhibit K-type spectra, but have exceptionally blue
colors; the $(J-H)\simeq-0.9$ Second Data Release 2MASS colors are
certainly in error.  Two of the unusual objects are extra-galactic:  one
object is a known Seyfert 1 galaxy at $z=0.33$ \citep{marziani96} and the
other is a known quasar at $z=2.62$ \citep{barkhouse01}.


	Table \ref{tab:unusual} lists the 34 unusual objects.  Column (1)
is our identifier.  Column (2) is our object classification, or the
classification found in the literature if available.  Column (3) is the
J2000 right ascension in hours, minutes, and seconds.  Column (4) is the
J2000 declination in degrees, arcminutes, and arcseconds.  Column (5) is
the $V$ magnitude. Column (6) is the (\vr) color.  Column (7) is the 2MASS
Second Data Release $J$ magnitude.  Columns (8) and (9) are the 2MASS
$(J-H)$ and $(H-K)$ colors.

\section{CONCLUSIONS} \label{sec:concl}

We have observed 764 blue stars as part of the on-going Century Survey
Galactic Halo project, a survey of the Milky Way halo and thick-disk
populations. The primary part of the sample has $V<16.5$ mag and
$(\vr)<0.25$ mag; they are contained within the $1^\circ\times64^\circ$
Century Survey photometric strip. The secondary part of the sample
has $J<15.0$ mag and $(J-H)<0.15$ mag, and lies within the 2MASS region
adjacent to the Century Survey. We have discussed our techniques for
measuring radial velocities, effective temperatures ($T_{\rm eff}$),
surface gravities ($\log{g}$), metallicities ([Fe/H]), spectral types, and
distances from our stellar spectra and colors.

One of our main goals is the identification of a {\it bona-fide} sample of
BHB stars.  We thus have devoted special special attention to BHB
classification.  We compare BHB samples selected by the methods of
\citet{kinman94}, \citet{wilhelm99a}, and \citet{clewley02}, and
identified a combined sample of 55 high-probability BHB stars. The BHB
stars comprise 54\% of a photometric sample selected on $(\vr)<0.10$ mag,
$13<V<16.5$ mag. Similarly, BHB stars comprise 32\% of a photometric
sample selected on $(J-H)<0.10$ mag, $12<J<15$ mag from the 2MASS second
data release photometry.

The distances and radial velocities of our sample fit in the standard
picture.  The late F-type stars are nearby dwarfs in the thin and thick
disk.  The earlier-type stars, with their wide range of metallicity and
velocity, are largely thick-disk and halo stars.  The BHB stars range over
distances $2<z<13$ kpc and exhibit a line-of-sight velocity dispersion
with respect to the local standard of rest of $\sigma = 97$ km s$^{-1}$,
consistent with a halo population.

In addition, we find 34 unusual objects in our blue-selected sample.  One
of the unusual objects is a late B star 0.8 kpc above the Galactic plane,
and another is a DZ7 white dwarf. We also find three carbon-enhanced,
mildly metal-poor stars.

Next, we plan to carry out a kinematic analysis of the Century Survey
Galactic Halo Project sample, making use, where available, of recently
determined proper motions.  We plan to measure the systematic motions of
the stars with depth and location along the survey strip.  We will look
for stars moving together in star streams, and test the possible reality
of any detected stream(s).

We are also in the process of extending our observations of the Century
Survey Galactic Halo Project.  Expanding the survey will allow us to
measure structure and systematic motions with higher confidence across a
greater arc of the sky.


\acknowledgements

We thank Perry Berlind and Mike Calkins for obtaining most of the 1.5m
spectra. We thank Stephen Warren and collaborators for making their Sersic
profile-fitting routine available to us. We thank John Norris for carrying
out the auto-correlation function calculations employed for some of our
metallicity determinations. This project makes use of data products from
the Two Micron All Sky Survey, which is a joint project of the University
of Massachusetts and the Infrared Processing and Analysis
Center/California Institute of Technology, funded by NASA and the NSF.
This research also makes use of the SIMBAD database, operated at CDS,
Strasbourg, France. TCB acknowledges partial support for this work from
grants AST 00-98508 and AST 00-98549, awarded by the National Science
Foundation. CAP acknowledges partial support for this work from grant
AST-0086321, awarded by the National Science Foundation.

 \clearpage
\bibliographystyle{apj}
\bibliography{ReferencesHS}

 \clearpage

\begin{deluxetable}{lcccc}	
 \tablewidth{0pt}
\tablecaption{Observations\label{tab:obs}}
\tablehead{	\colhead{Observed}  	& \colhead{N} &
		\colhead{Mag limit}	& \colhead{Color Selection} &
		\colhead{Location (B1950)}
}
	\startdata
Spring 2001 & 367 & $V<15.5$ & $(\vr)<0.30$ & 
	$10\fh5<\alpha<13\fh5,~ 29^\circ<\delta<30^\circ$ \\
Spring 2002 & 286 & $V<16.5$ & $(\vr)<0.25$ & 
	$~8\fh5<\alpha<13\fh5,~ 29^\circ<\delta<30^\circ$ \\
	& & ~~~1-in-10 & $0.25<(\vr)<0.30$ & ~~\arcsec \\
Spring 2002 & 111 & $J<15.0$ & $(J-H)<0.15$ &
	$~8\fh5 < \alpha < 11\fh25,~ 28^\circ < \delta < 29^\circ$ \\
& & \arcsec & ~~\arcsec & 
	$11\fh25 < \alpha < 13\fh5,~ 30^\circ < \delta < 31^\circ$ \\

	\enddata
\end{deluxetable}

\begin{deluxetable}{llcc}		
\tablewidth{0pt}
\tablecaption{Spectral Line Indices\label{tab:lines}}
\tablehead{
	\colhead{Spectral Line Index\tablenotemark{a}} &
	\colhead{Spectral Type\tablenotemark{b}} & 
	\colhead{Valid Range\tablenotemark{c}} & 
	\colhead{RMS\tablenotemark{d}}
}
	\startdata
	Ca {\small II} 3933 \AA & 
$=21.4 + 26.3$~Ca {\small II} & $22 < {\rm st} < 40$ & $\pm1.1$ \\
	H$_{\rm sum}=$ H 3889 \AA + H 4101 \AA &
$=42.2 - 6.98$~H$_{\rm sum}$ & $22 < {\rm st} < 39$ & $\pm1.4$ \\
	~~~~~~~+ H 4340 \AA + H 4861 \AA\ &
$=5.23 + 5.97$~H$_{\rm sum}$ & $~9 < {\rm st} < 22$ & $\pm1.4$ \\
	CN 3860 \AA & $=37.2 + 24.2$~CN & $35 < {\rm st} < 43$ & $\pm1.5$ \\
& $=7.58 - 79.0$~CN & $~9 < {\rm st} < 17$ & $\pm1.5$ \\
	CH 4305 \AA & $=28.6 + 43.9$~CH & $33 < {\rm st} < 42$ & $\pm2.0$ \\
& $=16.5 + 93.6$~CH & $~1 < {\rm st} < 26$ & $\pm2.0$ \\
	Mg {\small I} 5175 \AA &
$=29.1 + 90.1$~Mg {\small I} & $34 < {\rm st} < 52$ & $\pm2.3$ \\
	Fe$_{\rm sum}=$ Fe 4383 \AA + Fe 4531 \AA &
$=23.6 + 73.0$~Fe$_{\rm sum}$ & $25 < {\rm st} < 48$ & $\pm2.3$ \\
	~~~~~~~~+ Fe 5015 \AA + Fe 5270 \AA & & & \\
\enddata
	\tablenotetext{a}{The line indices (e.g.\ Ca {\small II}) are flux
ratios, normalized by bandwidth, of the flux in a stellar spectral line
band divided by the flux in nearby continuum side bands. See
\citet{oconnell73} and \citet{worthey94} for the band definitions. }
	\tablenotetext{b}{Fits based on luminosity class V stellar spectra
from \citet{jacoby84}.}
	\tablenotetext{c}{Spectral types are quantified as B0 = 10, A0 =
20, F0 = 30, etc.}
	\tablenotetext{d}{RMS residual of the least-squares fit, used as 
the weight for the spectral classification.}
\end{deluxetable}

\begin{deluxetable}{lcccc}          
\singlespace
\tablewidth{0pt}
\tablecaption{Comparison with \citet{kinman94}\tablenotemark{a}\label{tab:kinman}}
\tablecolumns{5}
\tablehead{
        \colhead{ID} & \colhead{Brown et al.} & \colhead{} &
	\colhead{Kinman et al.} & \colhead{} \\
	\colhead{} & \colhead{BJA$_0$} & \colhead{$\lambda_{0.5}$} &
	\colhead{BJA$_0$} & \colhead{$\lambda_{0.5}$} \\
}
        \startdata
HD~60778   	& 1.52 & 27.1 & 1.37 & 60.1 \\
HD~64488   	& 1.28 & 47.9 & 1.27 & 64.5 \\
HD~74721   	& 1.44 & 42.1 & 1.33 & 60.2 \\
HD~86986   	& 1.44 & 39.3 & 1.33 & 57.8 \\
HD~105805   	& 1.10 & 51.6 & 1.07 & 71.1 \\
	\enddata
	\tablenotetext{a}{[The complete version of this table is in the
electronic edition of the Journal.  The printed edition contains only a
sample.]}
	\end{deluxetable}

\begin{deluxetable}{cccccccccccccc}	
\tablewidth{0pt}
 \rotate
\tablecaption{The BHB Selection\tablenotemark{a}\label{tab:bhbaa}}
\tablehead{
	\colhead{ID} & \colhead{Type} & \colhead{BJA$_0$} & \colhead{a)} &
	\colhead{$T_{\rm eff}$} & \colhead{$\log{g}$} & \colhead{b)} &
	\colhead{$BV0$} & \colhead{$D_{0.15}$} & \colhead{c)} &
	\colhead{$c$} & \colhead{$b$} & \colhead{d)} & \colhead{Tot} \\
	& & mag & & K & cm s$^{-2}$ & & mag & \AA & & & \AA & &
}
	\startdata
CHSS  33 & $21.6\pm1.2$ & $0.81\pm0.02$ & 0 & $9743\pm200$ & $3.9\pm0.3$ & 1 & $      -0.02\pm0.04$ & $31.4\pm0.5$ & 1 & $0.96\pm0.03$ & $ 9.0\pm0.2$ & 1 & 3 \\
CHSS  40 & $20.3\pm1.2$ & $1.06\pm0.02$ & 1 & $8524\pm200$ & $3.4\pm0.3$ & 1 & $\phm{-}0.13\pm0.04$ & $32.1\pm0.5$ & 0 & $0.86\pm0.02$ & $ 8.6\pm0.1$ & 1 & 3 \\
CHSS  45 & $21.2\pm1.2$ & $1.10\pm0.02$ & 1 & $8450\pm200$ & $3.5\pm0.3$ & 1 & $\phm{-}0.11\pm0.04$ & $33.4\pm0.5$ & 0 & $0.92\pm0.02$ & $ 9.3\pm0.1$ & 1 & 3 \\
CHSS  56 & $21.8\pm1.2$ & $1.14\pm0.02$ & 1 & $8827\pm200$ & $3.5\pm0.3$ & 1 & $\phm{-}0.08\pm0.04$ & $32.4\pm0.5$ & 1 & $0.95\pm0.02$ & $ 9.2\pm0.2$ & 1 & 4 \\
CHSS  63 & $21.2\pm1.2$ & $1.10\pm0.02$ & 1 & $9221\pm200$ & $3.5\pm0.3$ & 1 & $\phm{-}0.09\pm0.04$ & $29.1\pm0.4$ & 1 & $1.09\pm0.03$ & $ 8.9\pm0.1$ & 0 & 3 \\
	\enddata
	\tablenotetext{a}{[The complete version of this table is in the
electronic edition of the Journal.  The printed edition contains only a
sample.]}
\end{deluxetable}

\begin{deluxetable}{lcccccccccccccc}	
\tablewidth{0pt}
 \renewcommand{\arraystretch}{.6}
\rotate
\tablecaption{The BHB Standards\label{tab:bhba}}
\tablehead{
	\colhead{ID} & \colhead{Type} & \colhead{BJA$_0$} & \colhead{a)} &
	\colhead{$T_{\rm eff}$} & \colhead{$\log{g}$} & \colhead{b)} &
	\colhead{$BV0$} & \colhead{$D_{0.15}$} & \colhead{c)} &
	\colhead{$c$} & \colhead{$b$} & \colhead{d)} & \colhead{Tot} &
	\colhead{Class} \\
	& & mag & & K & cm s$^{-2}$ & & mag & \AA & & & \AA & & & 
}
	\startdata
HD  60778    & $21.4\pm1.2$ & $1.52\pm0.01$ & 1 & $7754\pm200$ & $3.4\pm0.3$ & 1 & $\phm{-}0.10\pm0.04$ & $32.7\pm0.5$ & 1 & $0.88\pm0.01$ & $ 8.9\pm0.1$ & 1 & 4 & BHB \\
HD  64488    & $22.3\pm1.2$ & $1.28\pm0.01$ & 1 & $8985\pm200$ & $3.6\pm0.3$ & 1 & $      -0.02\pm0.04$ & $31.4\pm0.4$ & 1 & $0.94\pm0.01$ & $ 8.9\pm0.1$ & 1 & 4 & A$_{\rm rot}$ \\
HD  74721    & $21.8\pm1.2$ & $1.44\pm0.01$ & 1 & $8637\pm200$ & $3.4\pm0.3$ & 1 & $\phm{-}0.06\pm0.04$ & $31.8\pm0.5$ & 1 & $0.97\pm0.02$ & $ 9.1\pm0.1$ & 1 & 4 & BHB \\
HD  86986    & $21.0\pm1.2$ & $1.44\pm0.01$ & 1 & $7743\pm200$ & $3.1\pm0.3$ & 1 & $\phm{-}0.12\pm0.04$ & $32.5\pm0.5$ & 0 & $0.86\pm0.01$ & $ 8.7\pm0.0$ & 1 & 3 & BHB \\
HD 105805    & $23.8\pm1.2$ & $1.10\pm0.01$ & 0 & $7934\pm200$ & $3.9\pm0.3$ & 0 & $\phm{-}0.13\pm0.04$ & $37.5\pm0.5$ & 0 & $0.81\pm0.00$ & $ 9.6\pm0.0$ & 0 & 0 & A \\
HD 107131    & $26.1\pm1.4$ & $1.01\pm0.01$ & 0 & $7754\pm152$ & $3.8\pm0.3$ & 0 & $\phm{-}0.17\pm0.04$ & $34.2\pm0.5$ & 0 & $0.79\pm0.00$ & $ 8.6\pm0.0$ & 0 & 0 & A \\
HD 108382    & $23.8\pm1.3$ & $1.19\pm0.02$ & 0 & $8247\pm 58$ & $3.7\pm0.3$ & 1 & $\phm{-}0.16\pm0.04$ & $35.7\pm0.6$ & 0 & $0.87\pm0.01$ & $ 9.6\pm0.1$ & 0 & 1 & A \\
HD 109307    & $23.5\pm1.2$ & $1.06\pm0.02$ & 0 & $8153\pm200$ & $4.0\pm0.3$ & 0 & $\phm{-}0.12\pm0.04$ & $40.5\pm0.6$ & 0 & $0.82\pm0.00$ & $10.5\pm0.0$ & 0 & 0 & A \\
HD 109995    & $21.5\pm1.2$ & $1.32\pm0.01$ & 1 & $9152\pm200$ & $3.6\pm0.3$ & 1 & $\phm{-}0.08\pm0.04$ & $32.2\pm0.5$ & 1 & $0.93\pm0.01$ & $ 9.1\pm0.1$ & 1 & 4 & BHB \\
HD 161817    & $20.7\pm1.2$ & $1.26\pm0.01$ & 1 & $7129\pm200$ & $3.1\pm0.3$ & 1 & $\phm{-}0.16\pm0.04$ & $29.3\pm0.5$ & 0 & $0.84\pm0.01$ & $ 8.2\pm0.1$ & 1 & 3 & BHB \\
RR7  03      & $22.2\pm1.2$ & $1.04\pm0.03$ & 1 & $9743\pm200$ & $4.2\pm0.3$ & 0 & $\phm{-}0.05\pm0.04$ & $34.6\pm0.6$ & 1 & $0.94\pm0.03$ & $ 9.8\pm0.3$ & 0 & 2 & A \\
RR7  15      & $20.5\pm1.2$ & $1.21\pm0.01$ & 1 & $7386\pm200$ & $3.3\pm0.3$ & 1 & $\phm{-}0.18\pm0.04$ & $28.7\pm1.0$ & 0 & $0.76\pm0.02$ & $ 7.1\pm0.3$ & 1 & 3 & BHB \\
RR7  23      & $21.8\pm1.2$ & $1.30\pm0.02$ & 1 & $9157\pm200$ & $3.7\pm0.3$ & 1 & $\phm{-}0.06\pm0.04$ & $31.3\pm0.5$ & 1 & $1.00\pm0.04$ & $ 9.1\pm0.3$ & 0 & 3 & BHB \\
RR7  36      & $21.2\pm1.2$ & $1.22\pm0.03$ & 1 & $7602\pm200$ & $3.5\pm0.3$ & 1 & $      -0.03\pm0.04$ & $30.7\pm0.5$ & 1 & $0.84\pm0.02$ & $ 8.1\pm0.2$ & 1 & 4 & BHB \\
RR7  60      & $19.3\pm1.2$ & $1.22\pm0.02$ & 1 & $7443\pm200$ & $3.4\pm0.3$ & 1 & $\phm{-}0.15\pm0.04$ & $30.1\pm0.5$ & 0 & $0.79\pm0.02$ & $ 7.7\pm0.1$ & 1 & 3 & BHB \\
RR7  64      & $21.8\pm1.2$ & $1.33\pm0.01$ & 1 & $7951\pm1025$ & $3.5\pm0.3$ & 1 & $\phm{-}0.22\pm0.04$ & $33.1\pm0.5$ & 0 & $0.88\pm0.01$ & $ 9.0\pm0.1$ & 1 & 3 & BHB \\
RR7  78      & $22.4\pm1.2$ & $1.08\pm0.01$ & 1 & $9563\pm200$ & $4.2\pm0.3$ & 0 & $\phm{-}0.07\pm0.04$ & $38.6\pm0.5$ & 1 & $0.93\pm0.00$ & $10.8\pm0.0$ & 0 & 2 & A \\
RR7  91      & $21.7\pm1.2$ & $1.20\pm0.07$ & 1 & $9608\pm200$ & $3.8\pm0.3$ & 1 & $\phm{-}0.06\pm0.04$ & $31.6\pm0.5$ & 1 & $0.97\pm0.02$ & $ 9.1\pm0.1$ & 1 & 4 & BHB \\
RR7 103      & $22.3\pm1.0$ & $1.06\pm0.02$ & 1 & $9361\pm200$ & $4.3\pm0.3$ & 0 & $\phm{-}0.10\pm0.04$ & $39.5\pm0.6$ & 0 & $0.92\pm0.01$ & $11.0\pm0.1$ & 0 & 1 & A \\
SA57 01      & $20.1\pm1.2$ & $1.25\pm0.03$ & 1 & $7056\pm200$ & $3.0\pm0.3$ & 1 & $\phm{-}0.21\pm0.04$ & $28.4\pm1.6$ & 0 & $0.74\pm0.04$ & $ 7.0\pm0.6$ & 1 & 3 & BHB \\
SA57 06      & $21.6\pm1.2$ & $1.45\pm0.02$ & 1 & $8313\pm200$ & $3.2\pm0.3$ & 1 & $\phm{-}0.11\pm0.04$ & $33.0\pm0.5$ & 0 & $0.92\pm0.01$ & $ 9.2\pm0.1$ & 1 & 3 & BHB \\
SA57 07      & $21.0\pm1.2$ & $1.38\pm0.01$ & 1 & $9385\pm200$ & $3.4\pm0.3$ & 1 & $      -0.02\pm0.04$ & $28.8\pm0.5$ & 1 & $1.05\pm0.01$ & $ 8.8\pm0.1$ & 0 & 3 & BHB? \\
SA57 17      & $21.1\pm1.2$ & $1.50\pm0.01$ & 1 & $8503\pm200$ & $3.1\pm0.3$ & 1 & $\phm{-}0.09\pm0.04$ & $32.5\pm0.5$ & 1 & $0.91\pm0.01$ & $ 9.0\pm0.1$ & 1 & 4 & BHB \\
SA57 49      & $20.8\pm1.2$ & $1.32\pm0.01$ & 1 & $9487\pm200$ & $3.5\pm0.3$ & 1 & $\phm{-}0.06\pm0.04$ & $27.8\pm0.5$ & 1 & $1.06\pm0.02$ & $ 8.6\pm0.1$ & 0 & 3 & BHB? \\
SA57 80      & $21.5\pm1.2$ & $1.44\pm0.02$ & 1 & $8331\pm200$ & $3.2\pm0.3$ & 1 & $\phm{-}0.07\pm0.04$ & $32.3\pm0.5$ & 1 & $0.95\pm0.02$ & $ 9.2\pm0.1$ & 1 & 4 & BHB \\
M92   I - 10 & $21.3\pm1.2$ & $1.32\pm0.02$ & 1 & $9743\pm200$ & $3.6\pm0.3$ & 1 & $\phm{-}0.03\pm0.04$ & $29.5\pm0.4$ & 1 & $1.00\pm0.02$ & $ 8.6\pm0.1$ & 1 & 4 & BHB \\
M92  II - 23 & $21.3\pm1.2$ & $1.34\pm0.02$ & 1 & $9384\pm200$ & $3.6\pm0.3$ & 1 & $\phm{-}0.12\pm0.04$ & $33.0\pm0.6$ & 0 & $0.94\pm0.03$ & $ 9.3\pm0.2$ & 1 & 3 & BHB \\
M92  IV - 27 & $19.7\pm1.2$ & $1.26\pm0.03$ & 1 & $9373\pm200$ & $3.4\pm0.3$ & 1 & $\phm{-}0.11\pm0.04$ & $31.4\pm0.5$ & 0 & $0.82\pm0.02$ & $ 8.2\pm0.1$ & 1 & 3 & BHB \\
M92 XII - 01 & $17.7\pm1.2$ & $1.16\pm0.02$ & 1 & $7345\pm200$ & $3.4\pm0.3$ & 1 & $\phm{-}0.16\pm0.04$ & $26.0\pm1.3$ & 0 & $0.78\pm0.04$ & $ 6.9\pm0.5$ & 1 & 3 & BHB \\
M92 XII - 09 & $18.9\pm1.2$ & $1.31\pm0.02$ & 1 & $8699\pm200$ & $3.0\pm0.3$ & 1 & $\phm{-}0.19\pm0.04$ & $28.6\pm0.5$ & 0 & $0.80\pm0.02$ & $ 7.3\pm0.1$ & 1 & 3 & BHB \\
M92 XII - 10 & $20.3\pm1.2$ & $1.27\pm0.03$ & 1 & $9522\pm200$ & $3.7\pm0.3$ & 1 & $\phm{-}0.13\pm0.04$ & $30.6\pm0.5$ & 0 & $0.91\pm0.02$ & $ 8.5\pm0.1$ & 1 & 3 & BHB \\
M92   S - 20 & $21.2\pm1.2$ & $1.44\pm0.02$ & 1 & $8743\pm200$ & $3.3\pm0.3$ & 1 & $\phm{-}0.12\pm0.04$ & $32.6\pm0.5$ & 0 & $0.88\pm0.01$ & $ 8.9\pm0.1$ & 1 & 3 & BHB \\
M92   S - 24 & $20.9\pm1.2$ & $1.37\pm0.02$ & 1 & $9031\pm200$ & $3.5\pm0.3$ & 1 & $\phm{-}0.03\pm0.04$ & $32.8\pm0.6$ & 1 & $0.93\pm0.03$ & $ 9.2\pm0.2$ & 1 & 4 & BHB \\
	\enddata
\end{deluxetable}

\begin{deluxetable}{lcrcccccc}		
 \renewcommand{\arraystretch}{.6}
\tablewidth{0pt}
\tablecaption{Unusual Objects\label{tab:unusual}}
\tablecolumns{9}
\tablehead{
	\colhead{ID} & \colhead{Object} &
	\colhead{$\alpha_{\rm J2000}$} & 
	\colhead{$\delta_{\rm J2000}$} &
	\colhead{$V$} & \colhead{(\vr)} &
	\colhead{$J$} & \colhead{($J-H$)} & \colhead{($H-K$)} \\
	\colhead{} & \colhead{} & \colhead{h m s} & 
	\colhead{$^\circ ~ \arcmin ~\arcsec$} &
	\colhead{mag} & \colhead{mag} &
	\colhead{mag} & \colhead{mag} & \colhead{mag}
}
	\startdata
CHSS 167 & DA3    & 11:36:14.0 & 29:01:30 & $14.87\pm0.040$ & $      -0.15\pm0.06$ & $14.85\pm0.04$ & $\phm{-}0.65\pm0.06$ & $\phm{-}0.12\pm0.07$ \\
CHSS 583 & DA2    &  9:09:19.0 & 29:29:29 & $15.80\pm0.035$ & $      -0.14\pm0.05$ &        \nodata &              \nodata &              \nodata \\
CHSS 662 & DA6    &  9:33:41.1 & 29:11:24 & $15.96\pm0.041$ & $\phm{-}0.12\pm0.07$ &        \nodata &              \nodata &              \nodata \\
CHSS 653 & DA3    &  9:32:04.1 & 28:50:45 & $16.48\pm0.034$ & $      -0.14\pm0.05$ &        \nodata &              \nodata &              \nodata \\
CHSS 720 & DA1    & 10:02:22.5 & 29:27:55 & $16.41\pm0.031$ & $      -0.24\pm0.05$ &        \nodata &              \nodata &              \nodata \\
CHSS 887 & DZ7    & 13:10:13.4 & 29:43:59 & $15.91\pm0.034$ & $\phm{-}0.23\pm0.05$ & $14.52\pm0.04$ & $\phm{-}0.32\pm0.07$ & $      -0.05\pm0.12$ \\
CHSS  82 & BHB    & 10:59:28.0 & 29:25:09 & $13.54\pm0.038$ & $      -0.13\pm0.05$ & $13.80\pm0.03$ & $      -0.06\pm0.05$ & $      -0.08\pm0.07$ \\
CHSS 633 & B8.5   &  9:26:15.2 & 28:03:23 &       \nodata   &             \nodata  & $10.16\pm0.02$ & $      -0.09\pm0.03$ & $\phm{-}0.00\pm0.03$ \\
CHSS 711 & BHB    &  9:58:15.2 & 28:52:33 & $13.04\pm0.021$ & $      -0.07\pm0.03$ & $13.23\pm0.03$ & $      -0.07\pm0.05$ & $\phm{-}0.00\pm0.06$ \\
CHSS 748 & BHB    & 10:13:56.3 & 29:06:15 & $13.97\pm0.036$ & $      -0.07\pm0.05$ & $14.17\pm0.03$ & $      -0.15\pm0.07$ & $\phm{-}0.02\pm0.09$ \\
CHSS 114 & sdB    & 11:13:04.4 & 29:07:46 & $14.07\pm0.032$ & $      -0.18\pm0.05$ &        \nodata &              \nodata &              \nodata \\
CHSS 148 & sdB    & 11:28:29.3 & 29:15:04 & $15.05\pm0.027$ & $      -0.23\pm0.04$ &        \nodata &              \nodata &              \nodata \\
CHSS 409 & sdB    & 11:08:21.6 & 29:36:49 & $15.52\pm0.029$ & $      -0.20\pm0.04$ &        \nodata &              \nodata &              \nodata \\
CHSS 552 & sdA    &  8:57:02.0 & 29:10:48 & $15.52\pm0.024$ & $\phm{-}0.23\pm0.04$ & $14.60\pm0.03$ & $\phm{-}0.25\pm0.05$ & $      -0.01\pm0.07$ \\
CHSS 617 & sdO    &  9:23:13.4 & 29:26:58 & $14.69\pm0.033$ & $      -0.23\pm0.05$ &        \nodata &              \nodata &              \nodata \\
CHSS 678 & sdA    &  9:38:42.7 & 29:00:12 & $14.77\pm0.030$ & $\phm{-}0.24\pm0.05$ & $13.89\pm0.04$ & $\phm{-}0.26\pm0.06$ & $\phm{-}0.02\pm0.07$ \\
CHSS 773 & sdA    & 10:37:42.0 & 29:18:22 & $15.80\pm0.029$ & $\phm{-}0.22\pm0.05$ & $14.85\pm0.04$ & $\phm{-}0.20\pm0.07$ & $      -0.12\pm0.10$ \\
CHSS 800 & sdB    & 11:06:50.5 & 29:35:32 & $15.74\pm0.033$ & $      -0.18\pm0.05$ &        \nodata &              \nodata &              \nodata \\
CHSS 808 & sdB    & 11:19:04.8 & 29:51:53 &       \nodata   &             \nodata  & $14.89\pm0.04$ & $      -0.12\pm0.09$ & $\phm{-}0.05\pm0.16$ \\
CHSS 846 & sdA    & 12:11:21.0 & 29:19:28 & $16.18\pm0.028$ & $\phm{-}0.17\pm0.04$ &        \nodata &              \nodata &              \nodata \\
CHSS 862 & sdA    & 12:35:17.1 & 29:02:09 & $16.12\pm0.024$ & $\phm{-}0.24\pm0.04$ & $15.18\pm0.05$ & $\phm{-}0.07\pm0.10$ & $\phm{-}0.20\pm0.13$ \\
CHSS 197 & carbon & 11:55:09.1 & 28:58:11 & $14.90\pm0.035$ & $\phm{-}0.30\pm0.05$ & $13.89\pm0.03$ & $\phm{-}0.23\pm0.05$ & $\phm{-}0.01\pm0.06$ \\
CHSS 420 & carbon & 12:17:38.0 & 29:18:30 & $15.53\pm0.032$ & $\phm{-}0.27\pm0.05$ & $14.50\pm0.04$ & $\phm{-}0.30\pm0.07$ & $\phm{-}0.07\pm0.09$ \\
CHSS 538 & carbon &  8:53:24.1 & 28:44:46 &       \nodata   &             \nodata  & $14.86\pm0.05$ & $\phm{-}0.15\pm0.09$ & $\phm{-}0.13\pm0.12$ \\
CHSS 178 & compos & 11:43:37.2 & 28:56:00 & $14.45\pm0.023$ & $\phm{-}0.12\pm0.04$ & $13.98\pm0.03$ & $\phm{-}0.15\pm0.05$ & $\phm{-}0.02\pm0.06$ \\
CHSS 424 & compos & 12:40:27.6 & 28:54:07 & $15.66\pm0.029$ & $\phm{-}0.25\pm0.05$ & $14.65\pm0.05$ & $\phm{-}0.12\pm0.07$ & $\phm{-}0.20\pm0.09$ \\
CHSS 517 & K      &  8:42:25.1 & 29:13:36 & $16.22\pm0.041$ & $      -1.01\pm0.06$ &        \nodata &              \nodata &              \nodata \\
CHSS 591 & K      &  9:11:31.7 & 28:44:40 &       \nodata   &             \nodata  & $12.62\pm0.02$ & $      -0.89\pm0.14$ & $\phm{-}0.66\pm0.20$ \\
CHSS 615 & K      &  9:22:08.0 & 27:54:04 &       \nodata   &             \nodata  & $13.68\pm0.04$ & $      -0.95\pm0.08$ & $\phm{-}0.41\pm0.16$ \\
CHSS 750 & K      & 10:16:46.9 & 28:11:05 &       \nodata   &             \nodata  & $14.54\pm0.04$ & $      -0.01\pm0.07$ & $\phm{-}0.77\pm0.15$ \\
CHSS 802 & K      & 11:10:28.1 & 28:36:08 &       \nodata   &             \nodata  & $14.67\pm0.04$ & $      -0.49\pm0.09$ & $\phm{-}0.54\pm0.16$ \\
CHSS 807 & K      & 11:18:22.9 & 30:29:42 &       \nodata   &             \nodata  & $13.30\pm0.03$ & $      -0.84\pm0.14$ & $\phm{-}0.36\pm0.22$ \\
CHSS 724 & Seyfert& 10:04:02.6 & 28:55:35 & $15.84\pm0.029$ & $\phm{-}0.25\pm0.04$ &        \nodata &              \nodata &              \nodata \\
CHSS 744 & Quasar & 10:11:55.6 & 29:41:42 & $16.12\pm0.027$ & $\phm{-}0.16\pm0.04$ &        \nodata &              \nodata &              \nodata \\
	\enddata
 \end{deluxetable}

\begin{deluxetable}{lrcccccccc}		
 \tabletypesize \small
\tablewidth{0pt}
\tablecaption{Photometry\tablenotemark{a}\label{tab:dat1}}
\tablecolumns{10}
\tablehead{
	\colhead{ID} & 
	\colhead{$\alpha_{\rm J2000}$} & 
	\colhead{$\delta_{\rm J2000}$} &
	\colhead{$V$} & \colhead{(\vr)} &
	\colhead{$BV0$} & \colhead{$E(\bv)$} &
	\colhead{$J$} & \colhead{($J-H$)} & \colhead{($H-K$)} \\
	\colhead{} & \colhead{h m s} & 
	\colhead{$^\circ ~ \arcmin ~\arcsec$} &
	\colhead{mag} & \colhead{mag} &
	\colhead{mag} & \colhead{mag} &
	\colhead{mag} & \colhead{mag} & \colhead{mag} \\
}
	\startdata
CHSS 1 & 10:30:13.3 & 29:20:25 & $13.74\pm0.027$ & $\phm{-}0.29\pm0.04$ & $\phm{-}0.50$ & 0.02 & $12.71\pm0.03$ & $\phm{-}0.30\pm0.04$ & $\phm{-}0.05\pm0.04$ \\
CHSS 2 & 10:31:03.4 & 29:24:39 & $13.17\pm0.027$ & $\phm{-}0.28\pm0.04$ & $\phm{-}0.48$ & 0.02 & $12.10\pm0.02$ & $\phm{-}0.31\pm0.03$ & $\phm{-}0.04\pm0.03$ \\
CHSS 3 & 10:31:11.6 & 29:16:23 & $14.67\pm0.027$ & $\phm{-}0.27\pm0.04$ & $\phm{-}0.47$ & 0.02 & $13.66\pm0.03$ & $\phm{-}0.27\pm0.04$ & $\phm{-}0.06\pm0.04$ \\
CHSS 4 & 10:31:19.9 & 29:22:23 & $14.85\pm0.027$ & $\phm{-}0.28\pm0.04$ & $\phm{-}0.47$ & 0.02 & $13.87\pm0.03$ & $\phm{-}0.29\pm0.04$ & $\phm{-}0.04\pm0.05$ \\
CHSS 5 & 10:31:23.5 & 29:35:17 & $13.72\pm0.037$ & $\phm{-}0.30\pm0.06$ & $\phm{-}0.47$ & 0.02 & $12.71\pm0.03$ & $\phm{-}0.25\pm0.04$ & $\phm{-}0.06\pm0.04$ \\
	\enddata
 \tablenotetext{a}{[The complete version of this table is in the
electronic edition of the Journal.  The printed edition contains only a
sample.]}
 \end{deluxetable}

\begin{deluxetable}{lcccccl}		
\tablewidth{0pt}
\tablecaption{Spectroscopic Measurements\tablenotemark{a}\label{tab:dat2}}
\tablecolumns{7}
\tablehead{
	\colhead{ID} & 
	\colhead{KP} & 
	\colhead{HP2} &
	\colhead{HG2} & 
	\colhead{GP} &
	\colhead{Type} & 
	\colhead{V$_{\rm radial}$} \\
	\colhead{} & \colhead{} & \colhead{} & \colhead{} & \colhead{} & 
	\colhead{} & \colhead{km s$^{-1}$}
}
	\startdata
CHSS   1 & $\phm{1}8.14$ & $\phm{1}3.03$ & $\phm{1}3.18$ & 3.58 & $36.4\pm1.5$ & $\phm{-1}43.9\pm 9.7$ \\
CHSS   2 & $\phm{1}7.96$ & $\phm{1}3.42$ & $\phm{1}3.39$ & 3.32 & $36.6\pm1.6$ & $\phm{ }-18.9\pm 9.5$ \\
CHSS   3 & $\phm{1}7.81$ & $\phm{1}3.37$ & $\phm{1}3.84$ & 2.51 & $35.8\pm1.5$ & $\phm{-1}13.9\pm10.3$ \\
CHSS   4 & $\phm{1}7.05$ & $\phm{1}3.35$ & $\phm{1}3.43$ & 3.51 & $35.7\pm1.4$ & $\phm{ }-62.7\pm10.6$ \\
CHSS   5 & $\phm{1}8.42$ & $\phm{1}3.31$ & $\phm{1}3.21$ & 3.68 & $37.3\pm1.5$ & $\phm{-1}42.3\pm 9.5$ \\
	\enddata
 \tablenotetext{a}{[The complete version of this table is in the
electronic edition of the Journal.  The printed edition contains only a
sample.]}
 \end{deluxetable}

\begin{deluxetable}{lcccccccc}		
\tablewidth{0pt}
\tablecaption{Stellar Parameters\tablenotemark{a}\label{tab:dat3}}
\tablehead{
	\colhead{ID} & 
	\colhead{$T_{\rm eff}$} & 
	\colhead{$\log{g}$} &
	\colhead{[Fe/H]$_{\rm GA}$} &
	\colhead{[Fe/H]$_{\rm KP}$} &
	\colhead{[Fe/H]$_{\rm EC}$} &
	\colhead{[Fe/H]$_{\rm final}$} &
	\colhead{Dist} &
	\colhead{$M_V$} \\
	& K & cm s$^{-2}$ & & & & & kpc & mag
}
	\startdata
CHSS   1\tablenotemark{b} & 6013 & 4.25 &       $-0.51$ &       $-0.87$ &       $-0.77$ &       $-0.69$\phm{~:} & $\phm{1}0.72$ & $\phm{-}4.37$ \\
CHSS   2\tablenotemark{b} & 6088 & 4.56 &       $-0.39$ &       $-0.78$ &       $-0.87$ &       $-0.59$\phm{~:} & $\phm{1}0.67$ & $\phm{-}3.97$ \\
CHSS   3\tablenotemark{b} & 6182 & 4.48 &       $-0.44$ &       $-0.57$ &       $-0.84$ &       $-0.51$\phm{~:} & $\phm{1}1.26$ & $\phm{-}4.10$ \\
CHSS   4\tablenotemark{b} & 6095 & 4.45 &       $-0.82$ &       $-0.80$ &       $-0.75$ &       $-0.81$\phm{~:} & $\phm{1}1.26$ & $\phm{-}4.29$ \\
CHSS   5\tablenotemark{b} & 6148 & 4.70 &       $-0.45$ &       $-0.34$ &       $-0.19$ &       $-0.40$\phm{~:} & $\phm{1}0.86$ & $\phm{-}3.97$ \\
	\enddata
 \tablenotetext{a}{[The complete version of this table is in the
electronic edition of the Journal.  The printed edition contains only a
sample.]}
 \tablenotetext{b}{Spectrum observed in non-photometric conditions.}
 \end{deluxetable}

\appendix
\section{DATA TABLES}

Tables \ref{tab:dat1}, \ref{tab:dat2}, and \ref{tab:dat3} present the
photometric and spectroscopic measurements for our blue star sample.  The
tables contain 737 stars.  We exclude 27 unusual objects -- the white
dwarfs, subdwarfs, K stars, composite spectra, and QSOs -- but retain the
four B stars and the three carbon-enhanced stars. The complete version of
Tables are in the electronic edition of the Journal.  The printed edition
contains only a sample.

Table \ref{tab:dat1} summarizes the photometry. Column (1) is our
identifier.  The designation CHSS stands for Century Halo Star Survey and
is chosen to be unique from previous surveys.  Column (2) is the J2000
right ascension in hours, minutes, and seconds.  Column (3) is the J2000
declination in degrees, arcminutes, and arcseconds.  Column (4) is the $V$
magnitude.  Column (5) is the (\vr) color. Column (6) is the $BV0$ color
predicted from Balmer line strengths.  Column (7) is the $E(\bv)$
reddening value from \citet{schlegel98}, reduced according to our distance
estimate.  Column (8) is the 2MASS Second Data Release $J$ magnitude.  
Columns (9) and (10) are the 2MASS $(J-H)$ and $(H-K)$ colors.

Table \ref{tab:dat2} summarizes the spectral measurements:  the line
strengths, the estimated spectral type, and the radial velocity.  Column
(1) is our identifier. Column (2) is the KP (Ca {\small II}) index.  
Column (3) is the HP2 (H$\delta$) index. Column (4) is the HG2
(H$\gamma$) index.  Column (5) is the GP (G-band) index.  Column (6) is
the spectral type, where B0=10, A0=20, F0=30, and so forth.  Column (7) is
the heliocentric radial velocity in km s$^{-1}$.

Table \ref{tab:dat3} summarizes the stellar parameters. Column (1) is our
identifier.  Column (2) is the effective temperature in K.  Column (3) is
the surface gravity in cm s$^{-2}$.  Columns (4) through (6) are the
metallicities derived based on the genetic algorithm (GA), the KP index
(KP), and the equivalent widths/chi-square spectral match procedure (EC).  
Metallicities are given as the logarithmic [Fe/H] ratio relative to the
Sun.  Column (7) is the average [Fe/H] we take as the final value, as
described in \S \ref{sec:pars}.  Column (8) is the estimated distance in
kpc, as described in \S \ref{sec:dist}.  Objects with spectra obtained in
non-photometric conditions are marked and have increased uncertainty in
their distance estimates.  Column (9) is the absolute $M_V$ magnitude
corrected for reddening, given the estimated distance.  When we do not
have $V$ photometry, we use the 2MASS $J$ magnitude and add the mean
$(V-J)$ color for the measured spectral type to estimate $M_V$.

\end{document}